\documentclass[preprint,authoryear,12pt]{elsarticle}
\usepackage{amssymb,amsfonts,amsthm}
\usepackage{geometry}
\usepackage{graphics,subfigure}
\usepackage{hyperref}
\usepackage{pifont}
\usepackage{natbib}
\usepackage[fleqn]{amsmath}
\usepackage{color}
\usepackage{lineno}
\usepackage{bm}
\numberwithin{equation}{section}
\begin{document}
\title{Inflation-induced aneurysm formation and evolution in graded cylindrical tubes of arbitrary thickness}

\author[add1,add2]{Yang Liu\corref{cor1}}
\ead{tracy\underline{ }liu@tju.edu.cn}
\author[add1]{Liu Yang}
\author[add1,add2]{Yu-Xin Xie\corref{cor1}}
\ead{xyx@tju.edu.cn}

\address[add1]{Department of Mechanics, School of Mechanical Engineering, Tianjin University, Tianjin 300354, China}
\address[add2]{Tianjin Key Laboratory of Modern Engineering Mechanics, Tianjin 300354, China}

\cortext[cor1]{corresponding author}
\begin{abstract}
We study the initiation and evolution of aneurysmal morphology in a pressurized soft tube where the elastic modulus is non-uniform in the radial direction. The primary deformation prior to instability is characterized within the framework of nonlinear elasticity for a general material constitution and a generic modulus gradient. To unravel the influence of modulus gradient on aneurysm formation, we employ the incompressible Gent model and select three representative modulus gradients, including a linear, an exponential, and a sinusoidal function. In particular, the sinusoidal distribution can be used to model actual artery structure. In addition, two prototypical loading conditions are considered, namely, either the resultant axial force or the axial length can be fixed. Based on an explicit bifurcation condition in terms of the internal pressure and the resultant axial force for aneurysm formation or localized bulging, an exhaustive theoretical analysis on bulge initiation is carried out and the effect of geometric and material parameters and modulus gradient on the critical stretch generating localized bulging is revealed. It turns out that the modulus mismatch, as well as the position of maximum modulus, can dramatically affect the onset of localized bulging. Then we analytically elucidate the influence of modulus gradient on bulge propagation according to Maxwell's equal-area rule and conduct a finite element analysis of bulge evolution. To perform post-bifurcation analysis, a robust finite element model for graded Gent material is established in Abaqus by UHYPER subroutine coding. Interestingly, it is found that a sinusoidally distributed modulus has negligible influence on the critical stretch of bulge initiation, the deformation process of bugle growth, and the maximum size of a bulge. The current analysis not only can be used to qualitatively explain why a healthy human artery evolves into a sandwich structure where the intermediate layer is stiffest but also can provide useful insight into localized instabilities such as necking and beading in graded structures. 
\end{abstract}

\begin{keyword}
Graded tubes \sep Aneurysm formation \sep Bulge evolution  \sep Nonlinear elasticity \sep Parametric study  \sep Finite element analysis
\end{keyword}
\maketitle
\section{Introduction}
Inflation of a cylindrical rubber tube usually generates a prototypical localization instability, known as localized bulging, and such an interesting phenomenon was first recorded in detail and primarily analyzed by \cite{mallock}. With applications to bulge initiation and propagation in pressure vessels \citep{jam1984,ijms1988,book2007} and aneurysm formation in human arteries \citep{new2011,sa2018}, inflation of a cylindrical hyperelastic tube offers a pertinent paradigm to study the mechanism of localized instabilities \citep{ky1990,ky1991,prsa2021}. In addition, recent advances on inflated tubes and localized bulging can shed light on bifurcation in arteries affected by Marfan's syndrome \citep{mrc2009,mrc2010}, bulge prevention in energy harvesting devices \citep{re2013}, inflation-driven soft robots \citep{jin2021}, inflation of nematic elastomers \citep{he2020}, and bulge formation subject to additional effects of electric actuation \citep{lu2015}, swelling \citep{dm2017},  magnetic field \citep{ijss2018}, and plasticity \citep{takla2}. 

In general, the evolution of localized bulging is mainly composed of three typical periods, including bulge initiation, growth, and propagation \citep{wang2019}. Specifically, experimental evidence suggests that the diameter of the bulge will reach a maximum in the propagation stage, and a two-phase deformation can be observed \citep{ky1990}. On the theoretical side, the pioneering works on bifurcation analysis of inflated membrane tubes and thick tubes were conducted by \cite{ho79,ho791}. Later, \cite{fu2008} found that a zero mode, which is usually viewed as another uniform deformation, precisely generates localized bulging, and a bifurcation condition was derived for a membrane tube based on the dynamical systems theory. Since then, multiple fundamental problems were investigated, such as solution stability \citep{pearce2010,fuxie2010,ams2012}, influence of material constitution \citep{pearce2012}, imperfection sensitivity \citep{fuxie2012}, and dynamic inflation \citep{dynamic}. Within the framework of nonlinear elasticity, an explicit bifurcation condition for thick tubes was derived by \cite{fu2016}, and the accuracy of membrane assumption was examined. This bifurcation condition paves a convenient way to study localized bugling in rotating cylinders \citep{wang2017}, fiber-reinforced tubes \citep{wangfu2018}, and layered tubes \citep{liu2019,ye2019}. Furthermore, a comprehensive understanding of localized bulging can provide useful insight into localized necking \citep{eml2018}. We mention that the evolution of a bulge can only be handled in the nonlinear regime. To perform post-bifurcation analysis for inflated thick-walled cylinders, \cite{fead2013} presented a numerical procedure by virtue of nonlinear finite element models. Within the framework of nonlinear elasticity,  \cite{ye2020} carried out a weakly nonlinear analysis to deduce the amplitude equation for a bulge and discovered an interesting parametric domain where necking occurs instead of bulging. Recently, the analysis of localizations was further extended to situations where surface tension is involved \cite{wang2020,emery2021,emery20211,fu2021}. Recent papers have shown a growing interest in the community for such a paradigmatic localization problem.

In addition to the investigations mentioned above, some effort was dedicated to establishing a reduced model applicable to inspect bulge formation and propagation in cylindrical hyperelastic tubes. \cite{audoly2018} derived a one-dimensional model for the analysis of localized bulging in long tubes making use of nonlinear membrane theory and asymptotic expansion. A fundamental assumption is that localization can be viewed as a long-wavelength periodic mode. Subsequently, \cite{biggins2020} proposed an energy method in virtue of asymptotic expansions to track the nonlinear evolution of an inhomogeneous solution and further to identify phase separation of typical localization such as beading, bulging, and necking.

As mentioned earlier, localized bulging in inflated tubes can be viewed as an analogue to aneurysm formation in arteries \citep{sh2001,fu2012}. Accordingly, theoretical or numerical analysis was performed to supply mechanical insight into possible pathogenesis behind aneurysms, see \cite{ijes2014,fu2015,vn2017,vn2018} and the references therein. In particular, our previous analysis reveals a fact that an aneurysm is more prone to appear as the innermost layer of an artery stiffens \citep{ye2019}. This qualitatively interprets why the risk of aneurysm formation for a person becomes higher with aging \citep{fg2015}.    

Generally speaking, two loading conditions, i.e. either the resultant axial force or the axial length is specified, are adopted in experiments. The former was adopted in \cite{ky1990,ky1991,wang2019,prsa2021} while the latter was used in \cite{ijms2006,ijms2008,guo2016,wang2019,prsa2021}. On the one hand, the resultant axial force can be fixed by suspending a dead weight at one end while air comes into the tube from the other end. In this scenario, the curve of pressure versus volume ratio possesses an $N$-shape. Therefore, localized bulging is induced by a limit-point instability \citep{ijes1971,ijnm2007,wt2018} and the propagation pressure can be predicted by Maxwell's equal-area rule \citep{jam1984}. On the other hand, the fixed axial length can be acquired by imposing a pre-stretch on the tube and then fixing the total length. In this situation, the pressure-volume ratio curve may be monotonic. The effect of pre-stretch on pressurized thin tubes was examined by \cite{mao2014,hnijms}. Furthermore, human arteries usually suffer a fixed pre-stretch \textit{in vivo} \citep{bmm2014}. In this study, we also employ these two loading approaches.

It is pointed out that most existing literature on localized bulging is mainly concerned with a homogeneous or piecewise homogeneous material. Yet functionally graded materials (abbreviated by FGMs) with continuously varied physical or mechanical properties have unfolded many unusual functions such as high-temperature resistance and diminishing stress or strain concentration. This new type of material first emerged in the middle of the 1980s by mixing metal and ceramic phases in a manageable way. In doing so, it will acquire a desirable material property that distributes spatially. Afterwards, the mechanics of FGM has rapidly taken a center stage in solid mechanics \citep{ke2008,zhong2012,jha2013}. Especially, the popularity of 3D printing techniques substantially facilitates the fabrication of FGMs. Besides, human arteries chiefly consist of intima, media, and adventitia \citep{je2000,hgo}, which can be viewed as a sandwich structure and further a special graded structure. In most practical applications, we intend to prevent localized bulging or aneurysm formation, such as an Anaconda wave-energy extraction device \citep{re2013}. Naturally, improving the stiffness of a homogeneous tube will raise the critical pressure inducing localized bulging since pressure is proportional to the elastic modulus. However, an extremely stiff structure will lose the ability to deform subjected to a similar load and hence will degrade certain structural or physiological performance, particularly for arteries. Naturally, a compromised way is to partially stiffen. Then a well-imposed question arises.  Where is the optimal position to be strengthened, the innermost surface, the outermost surface, or the middle part? In fact, the question has been perfectly solved by nature. It is the intermediate layer that is hardest, and this is also the optimization result of artery evolution \citep{je2000,jctr2012}. But why? Previous studies have shown the potential effect of modulus gradient form on the stress distribution in functionally graded rubberlike cylinders and spheres \citep{mms2011} or pattern transition in growing graded tubular tissues \citep{liu2020}. It is therefore of fundamental significance to clarify the influence of material inhomogeneity, different modulus gradients as well as the maximum modulus mismatch on bulge initiation, growth, and propagation. This motivates the current study. Although deformation or bifurcation analysis was carried by \cite{bb2009} for inflated and everted circular cylinders comprised of a graded Mooney-Rivlin material and by \cite{chenwq2017} for pressurized graded cylindrical tubes, respectively, localization instability was still not addressed. In this study, we aim at furnishing a thorough analysis of localized bulging in graded hyperelastic tubes under the combined action of internal pressure and axial stretching and unveiling the mechanism behind structure optimization of arteries.

This paper is organized as follows. In Section 2, we characterize the primary deformation generated by inflation for a general material constitution and a generic modulus gradient and establish all required formulas used in the bifurcation condition within the framework of nonlinear elasticity. Section 3 demonstrates the numerical strategy for identifying bulge initiation and presents a detailed parametric study for the onset of localized bulging. A finite element model is constructed and validated in Section 4. A nonlinear analysis based on finite element analysis is performed in Section 5 and the bulge propagation is investigated by Maxwell's equal-area rule. Finally, some conclusions are given in Section 6.

\section{Primary deformation and bifurcation condition}
\begin{figure}[!htbp]
\centering\includegraphics[scale=0.8]{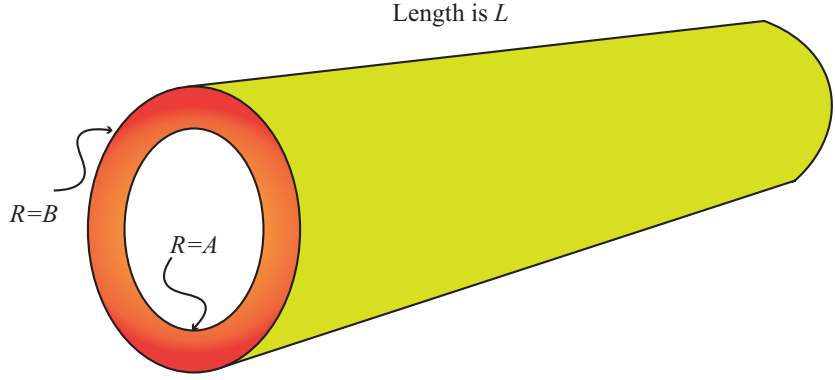}\caption{(Color online) A graded tube with inner radius A, outer radius B and length $L$ in its stress-free state.}\label{fig1}
\end{figure}
As shown in Figure \ref{fig1}, a cylindrical tube with inner radius $A$ and outer radius $B$ in its initial state is considered in this study. Since we focus on a graded soft tube, it is assumed that the elastic modulus of the tube is no longer uniformly distributed in the radial direction but is homogeneous in the axial direction. When such a graded tube is subjected to an internal pressure $P$ and an axial extension, a primary deformation is first generated, and the inner and outer radii become $a$ and $b$, respectively. Then we define the initial state as the reference configuration while the primary state the current configuration. Assuming that the tube is composed of incompressible hyperelastic material, we can use  $W(\mathbf{F})$ to represent the strain-energy function where $\mathbf{F}$ is the deformation gradient and is subjected to a constraint $\operatorname{det} \mathbf{F}=1$, which is the so-called incompressibility condition.

For the current problem, it is convenient to adopt cylindrical polar coordinates in both the reference and current configurations. By determining the centroid as the origin, the position coordinates of the same material point in the reference and current states are denoted by $(R, \Theta, Z)$ and $(r, \theta, z)$, respectively, and the tube occupies $A\leqslant R\leqslant B$ and $-L/2\leqslant Z\leqslant L/2$ in the reference state. Bearing in mind that the primary deformation is axisymmetric, we can write the deformation gradient as
\begin{equation}
\mathbf{F}=\lambda_1\bm e_r\otimes\bm e_r+\lambda_2\bm e_\theta\otimes\bm e_\theta+\lambda_3\bm e_z\otimes\bm e_z,\label{eq2_1}
\end{equation} 
where $\lambda_i$ $(i=1,2,3)$ is the principal stretch in $i$-direction and the subscripts 1, 2 and 3 denote the $r$-, $\theta$- and $z$-directions. Furthermore, the common orthonormal basis $\{\bm e_r, \bm e_\theta, \bm e_z\}$ is used in both states. In particular, the first two principal stretches are equal to
\begin{equation}
\lambda_1=\dfrac{\textrm{d} r}{\textrm{d} R},~~\lambda_2=\dfrac{r}{R},\label{eq2_2}
\end{equation}
and the third one $\lambda_3$ is a constant since the primary deformation is homogeneous in the $z$-direction. 

The strain-energy function can be written as a function of these principal stretches $W(\lambda_1,\lambda_2,\lambda_3)$. Accordingly, the non-zero components of Cauchy Stress tensor are given by 
\begin{align}
\sigma_{ii}=\lambda_iW_{,i}-p,~~\textrm{no summation}, \label{eq2_3}
\end{align}
where $p$ depicts the Lagrange multiplier enforcing the incompressibility condition. Meanwhile, we have defined that a comma means derivative with respect to the corresponding variable, for instance, $W_{,2}=\partial W/\partial \lambda_2$. 

It can be readily checked that the primary deformation is governed by the sole equilibrium equation: 
\begin{align}
\dfrac{\textrm{d} \sigma_{rr}}{\textrm{d} r}+\dfrac{\sigma_{rr}-\sigma_{\theta\theta}}{r}=0.\label{eq2_4}
\end{align}

In order to facilitate subsequent analysis, we define two parameters according to the incompressibility condition 
\begin{align}
\lambda=\lambda_2=\frac{r}{R},~\lambda_z=\lambda_3.\label{eq2_5}
\end{align}
Correspondingly, the first principal stretch is determined by $\lambda_1=1/(\lambda \lambda_z)$. In addition, the following relations can be derived :
\begin{align}
&r^2=\lambda_z^{-1}(R^2-A^2)+a^2,~~\theta=\Theta,~~z=\lambda_zZ.\label{eq2_6}
\end{align}

Next, we replace $\lambda_1$ in $W$ and employ a reduced strain-energy function $w$ which is determined by
\begin{align}
w(\lambda,\lambda_z)\equiv W(1/(\lambda \lambda_z),\lambda,\lambda_z).\label{eq2_7}
\end{align}
Subsequently, we find that
\begin{align}
w_{,1}=W_{,2}-\frac{1}{\lambda^2\lambda_z}W_{,1},~~w_{,2}=W_{,3}-\frac{1}{\lambda\lambda_z^2}W_{,1}.\label{eq2_8}
\end{align}

Substituting the stress components in (\ref{eq2_3}) into the governing equation (\ref{eq2_4}) and utilizing (\ref{eq2_8}) yield
\begin{align}
\sigma_{rr}=\int_{r_0}^{r}\frac{\lambda w_{,1}}{r}\textrm{d}r, \label{eq2_9}
\end{align}
where $r_0$ is a constant. Actually $w$ is dependent on $R$ for a graded material. It can be deduced from $(\ref{eq2_5})_1$ and $(\ref{eq2_6})_1$ that $R=A\sqrt{\lambda_z\lambda_a^2-1}/\sqrt{\lambda_z\lambda^2-1}$. Then we apply a variable exchange $\lambda=r/R$ and a new identity of the Cauchy stress $\sigma_{rr}$ is given by
\begin{equation}
\sigma_{rr}=\int_{\lambda_0}^\lambda\frac{w_{,1}}{1-\lambda^2\lambda_z}\textrm{d}\lambda,\label{eq2_10}
\end{equation}
where $\lambda_0$ is another constant to be determined later and we have replaced $R$ by $A\sqrt{\lambda_z\lambda_a^2-1}/\sqrt{\lambda_z\lambda^2-1}$ such that the integrand in (\ref{eq2_10}) is only a function of $\lambda$ (see also equation (\ref{eq2_15})). 

Before we proceed further, it is appropriate to introduce two hoop stretches at the inner surface and outer surface as follows
\begin{equation}
\lambda_a=\frac{a}{A},~~\lambda_b=\frac{b}{B},\label{eq2_11}
\end{equation}
and their relation can be obtained from (\ref{eq2_6})
\begin{equation}
\lambda_b=\sqrt{\frac{B^2-A^2+A^2\lambda_a^2\lambda_z}{B^2\lambda_z}}.\label{eq2_12}
\end{equation}

It is assumed that the outer surface of the tube is traction-free and the inner surface suffers a pressure $P$. Therefore, we obtain
\begin{equation}
\sigma_{rr}|_{\lambda=\lambda_b}=0,~~\sigma_{rr}|_{\lambda=\lambda_a}=-P.\label{eq2_13}
\end{equation}
It then follows from the boundary condition $(\ref{eq2_13})_{1}$ that the constant of integration $\lambda_0$ is identical to $\lambda_b$. In addition, the other one $(\ref{eq2_13})_{2}$ furnishes
\begin{equation}
P=\int_{\lambda_b}^{\lambda_a}\frac{w_{,1}}{1-\lambda^2\lambda_z}\textrm{d}\lambda.\label{eq2_14}
\end{equation}
Actually, the pressure $P$ can be regarded as a function of $\lambda_a$ and $\lambda_z$. Similarly, the resultant axial force $N$ in any cross-section is also dependent on  $(\lambda_a,\lambda_z)$ and holds the following form
\begin{align}
N=2\pi\int_a^b\sigma_{zz}r\textrm{d}r-P\pi a^2=\pi A^2(\lambda_a^2\lambda_z-1)\int_{\lambda_b}^{\lambda_a}\frac{2\lambda_z w_{,2}-\lambda w_{,1}}{(\lambda^2\lambda_z-1)^2}\lambda \textrm{d}\lambda.\label{eq2_15}
\end{align}

It should be mentioned that (\ref{eq2_14}) and (\ref{eq2_15}) are valid for both homogeneous and graded materials. For graded tubes, the reduced strain-energy function in these two expressions contains radius-dependent elastic modulus. Moreover, (\ref{eq2_14}) can be seen as a special case in \cite{chenwq2017} where both the inner and outer surfaces suffer pressures.

Currently, the primary deformation induced by inflation has been fully characterized by the above two equations for a graded hyperelastic tube of finite thickness. In practice, there are usually two different end conditions, i.e. either the axial force $N$ or the axial length $\lambda_zL$ is fixed. The former case can be attained by imposing an object of dead weight at one end and leaving the other end free. Correspondingly, (\ref{eq2_14}) and (\ref{eq2_15}) provide two algebraic equations for the two unknowns $\lambda_a$ and $\lambda_z$ and solving them for a given pressure and a given axial force yields the solution that characterizes the primary deformation. This kind of loading type has been used in the experimental investigations carried out by \cite{ky1990,ky1991,guo2016,wang2019, prsa2021}. In the other loading condition, the tube is first uniformly stretched and then the two ends are fastened.  As a result, the axial length or the axial stretch $\lambda_z$ is fixed. Furthermore, the axial force $N$ is not controlled and may increase, decrease or remain constant during the inflation process. We emphasize that this scenario corresponds to the $in$ $vivo$ situation for human arteries \citep{bmm2014}. Also, such an experimental setup was employed by \cite{ijms2006,ijms2008,wang2019,prsa2021}. Since $\lambda_z$ is prescribed, the internal pressure $P$ is only a function of $\lambda_a$, and the primary deformation is fully determined by equation (\ref{eq2_14}).

In an earlier study by \cite{chenwq2017}, various bifurcation behaviors were investigated based on the incremental theory for functionally graded tubes subjected to internal pressure, external pressure, and end compression, and periodic modes in the axial and hoop directions were analyzed. In this study, we consider that the graded tube suffers a combined action of internal pressure and an axial extension. Therefore, zero-mode may become preferred \citep{fu2008}, resulting in localized bulging in soft cylindrical tubes. On the one hand, if either the resultant axial force $N$ or the prescribed axial stretch $\lambda_z$ is greater enough, localized bulging can disappear \citep{liu2019}.   On the other hand, if the axial stretch $\lambda_z$ is less than a threshold value, global buckling can replace localized bulging \citep{lin2020,prsa2021}. In this paper, we intend to investigate localized bulging (aneurysm formation) in graded tubes and further to elucidate the influence of modulus gradient on bulge formation, growth, and propagation. To this end, we merely concentrate on the situation that localized bulging is preferred as a result of the primary bifurcation.

Bearing in mind that we have denoted the internal pressure $P$ and the resultant axial force $N$ as functions of the stretches $\lambda_a$ and $\lambda_z$, it is suitable to resort to an explicit bifurcation condition that the Jacobian of $P$ and $F$ in terms of variables $\lambda_a$ and $\lambda_z$ vanishes. This concise bifurcation condition was first proposed by \cite{fu2016} in the framework of finite elasticity. Later, its applicability to localized bulging in bilayer tubes was verified numerically using FE analysis in commercial software Abaqus \citep{liu2019,ye2019}. Recently, \cite{yu2021} offered analytical proof of this bifurcation condition with two relatively weak assumptions. The first one is that the strain-energy function is isotropic while the second one is that the deformations before bulge initiation and in the propagation stage are both axisymmetric. So, it also supports the validity of this bifurcation condition when applied to graded tubes. Writing explicitly, we obtain the bifurcation condition for localized bulging as follows
\begin{equation}
J(P,N)=\frac{\partial P}{\partial\lambda_a}\frac{\partial N}{\partial\lambda_z}-\frac{\partial P}{\partial \lambda_z}\frac{\partial N}{\partial\lambda_a}=0.\label{eq2_16}
\end{equation}

With the aid of (\ref{eq2_16}), the first bifurcation points for fixed axial force and fixed axial length can be identified, respectively, by the following equations
\begin{equation}
\left\{
\begin{aligned}
&\textrm{fixed axial force (loading type I):}~~~
J(P,N)=0,~~
N(\lambda_a,\lambda_z)=N_0,\\&
\textrm{fixed axial length (loading type II):}~~~
J(P,N)=0,~~
\lambda_z=\lambda_{z0},
\end{aligned}
\right.\label{eq2_17}
\end{equation}
where $N_0$ and $\lambda_{z0}$ are given constants. For convenience, we abbreviate loading type I (II) as LTI (LTII) from now on. Indeed, equation $(\ref{eq2_17})_1$ furnishes two critical stretch values $\lambda_a^c$ and $\lambda_z^c$ for LTI where localized bulging initiates, and equation $(\ref{eq2_17})_2$ determines one critical stretch value $\lambda_a^c$ for LTII. Subsequently, the critical pressures $P_{cr}$ for both loading types can be calculated from equation (\ref{eq2_14}).

Next, we suppose that the tube is composed of the incompressible Gent material in all illustrative examples, and the strain-energy function is given by
\begin{equation}
W=-\frac{\mu(R)}{2}J_m\textrm{ln}\bigg(1-\frac{\lambda_1^2+\lambda_2^2+\lambda_3^2-3}{J_m}\bigg),\label{eq2_18} 
\end{equation}
where $\mu(R)$ stands for the radius-dependent shear modulus for a graded tube and $J_m$ is a material parameter representing the maximum extensibility. In particular, we set $J_m=97.2$ which is typical for rubber \citep{gent}. Note that this model can well capture the bulge initiation, bulge growth and bulge propagation in finite element simulations \citep{liu2019,ye2019}, and these three distinctive stages were commonly seen in experiments \citep{wang2019}. 

\begin{figure}[!htbp]
\centering\includegraphics[scale=1.2]{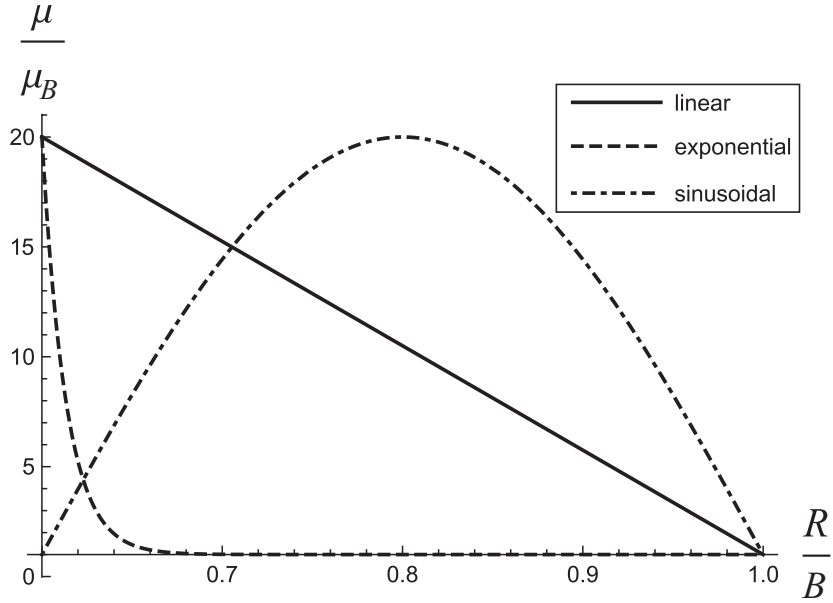}\caption{Three typical distributions of the shear modulus $\mu$ in the radial direction. The parameters are given by $\beta_1=\beta_2=20$, $A/B=0.8$, $\zeta=30$.}\label{fig2}
\end{figure}

As mentioned earlier, we are also concerned with how different modulus gradients affect the onset of localized bulging and bulge evolution. We emphasize that an accurate modulus distribution in a graded structure is difficult to describe. When fabricated polydimethylsiloxane (PDMS) material is exposed to UV radiation, Young's modulus may have an exponential distribution, resulting in a graded PDMS structure \citep{chen2018}. Generally speaking, the modulus distribution may be varied in different situations. A power law was adopted in \cite{bb2009} and a linear function was employed in \cite{chenwq2017}. In our previous study for growth-induced surface instabilities, it was found that different modulus distributions can alter pattern transition \citep{liu2020}. Moreover, it was shown in \cite{ye2019} that a stiffer inner layer makes the structure less stable. If a sandwich tube has a stiffer or softer core and the two faces share the same shear modulus, there must exist a competition among these three layers and may cause some interesting results. This intuitively motivates the investigation on a non-monotonic modulus distribution, which can further be used to model human arteries. As a result, we employ three modulus functions as follows
\begin{equation}
\left\{
\begin{aligned}
&\textrm{linear distribution:}~
\mu(R)=\mu_B\left[(\beta_1-1)\dfrac{R-B}{A-B}+1\right],\\&
\textrm{exponential distribution:}~\mu(R)=\mu_B\left[(\beta_1-1)\textrm{exp}\left(\zeta\dfrac{A-R}{B-A}\right)+1\right],\\&
\textrm{sinusoidal distribution:}~\mu(R)=\mu_B\left[(\beta_2-1)\sin\left(\dfrac{R-A}{B-A}\pi\right)+1\right],
\end{aligned}
\right.\label{eq2_19}
\end{equation}
where $\beta_1=\mu(A)/\mu_B$ signifies the ratio of the shear modulus of the inner surface to that of the outer surface and $\mu_B=\mu(B)$. The $\beta>1$ means that the inner surface is stiffer than the outer surface. In the sinusoidal distribution, we define $H=(B+A)/2$ and let $\beta_2=\mu(H)/\mu_B$. Moreover, the parameter $\zeta$ in $(\ref{eq2_19})_2$ refers to as the decay rate. In fact, letting $R=B$ in $(\ref{eq2_19})_1$ yields $\mu(B)=\mu_B\left((\beta_1-1)\textrm{exp}(-\zeta)+1\right)$. In order to satisfies $\mu(B)=\mu_B$, we must have $\zeta=+\infty$. In our illustrative examples, we specify $\zeta=30$ as an approximation such that exp$(-30)\approx 9.358\times10^{-14}$. Meanwhile, the effect of $\zeta$ is out of the scope of the current study. We then plot these three modulus distributions in Figure \ref{fig2} to depict the details.

In this section, we have derived the expressions of the pressure $P$ and the resultant axial force $N$ for the primary deformation of a graded soft tube. Meanwhile, a theoretical framework for determining the bulge initiation is established for a general material model and a general modulus gradient. In the next section, a detailed parametric analysis will be carried out based on the bifurcation condition (\ref{eq2_16}).

\section{Theoretical analysis of bulge initiation}
We have indicated that this study only focuses on the situation where localized mode is preferred. Accordingly, the prescribed axial force $N_{0}$ and the prescribed axial stretch $\lambda_{z0}$ in (\ref{eq2_17}) should lie in a proper domain. In the succeeding calculations, we specify $N_0=0$ and $\lambda_{z0}=1.5$ and suppose that the analysis can also shed light on qualitative effects of different parameters and modulus gradients for other $N_0$ and $\lambda_{z0}$. To facilitate further analysis, we introduce some dimensionless quantities as follows
\begin{align}
A^*=\dfrac{A}{B},~~N^*=\dfrac{N}{\mu_B B^2},~~\mu^*=\dfrac{\mu}{\mu_B},~~P^*=\dfrac{P}{\mu_B},~~w^*=\dfrac{w}{\mu_B}.
\end{align}
Correspondingly, the bifurcation condition for localized bulging is rewritten as $J(P^*, N^*)=0$.

In practice, analytical expressions of the dimensionless pressure $P^*$ and the dimensionless force $N^*$ can be derived for many homogeneous material models such as the neo-Hookean model, Mooney-Rivlin model, Ogden model, and Gent model. Accordingly, equation (\ref{eq2_17}) contains two sets of nonlinear algebraic equations, and solving them numerically can lead to the critical stretch $\lambda_a^c$. However, when the shear modulus becomes position-dependent in the Gent model, explicit formulas for $P^*$ and $N^*$ may be difficult to acquire for a complex modulus distribution, for instance, an exponential one. Therefore, we need to employ a numerical integration scheme. To construct a general computation framework for identifying the first bifurcation point, we will introduce our solution procedure first and then illustrate the bifurcation results.

\subsection{Solution strategy}
Without loss of generality, we apply the exponential modulus to exhibit the solution procedure, and all symbolic calculations are carried out in $Mathematica$ \citep{math2019}. This selection gives rise to that exponential functions are involved in the integrands of (\ref{eq2_14}) and (\ref{eq2_15}). So explicit primitive functions of these two integrations become difficult to obtain. We shall resort to the built-in command $NIntegrate$ in $Mathematica$. In the case of $N_0=0$, the two unknowns $\lambda_a$ and $\lambda_z$ can be identified simultaneously according to $(\ref{eq2_17})_1$. In order to generate effective numerical scheme, we define some necessary quantities as follows
\begin{equation}
\left\{
\begin{aligned}
&P_1(\lambda,\lambda_a,\lambda_z)=\dfrac{w^*_{,1}}{1-\lambda^2\lambda_z},\\&
N_1(\lambda_a,\lambda_z)=\pi(A^*)^2(\lambda^2_a\lambda_z-1),~~
N_2(\lambda,\lambda_a,\lambda_z)=\lambda\dfrac{2\lambda_zw^*_{,2}-\lambda w^*_{,1}}{(\lambda^2\lambda_z-1)^2}.
\end{aligned}
\right.\label{eq3_2}
\end{equation}
By use of the above new notations, we rewrite $P^*$ and $N^*$ as
\begin{align}
P^*=\int_{\lambda_b}^{\lambda_a}P_1(\lambda,\lambda_a,\lambda_z)\textrm{d}\lambda,~~N^*=N_1(\lambda_a,\lambda_z)\int_{\lambda_b}^{\lambda_a}N_2(\lambda,\lambda_a,\lambda_z)\textrm{d}\lambda.\label{eq3_3}
\end{align}

It is known that the bifurcation condition $J(P^*,N^*)=0$ actually describes a curve in the $(\lambda_a,\lambda_z)$-plane. Similarly, the loading condition $N^*=0$ provides another curve in the same plane. So, the intersection point corresponds to the first bifurcation point where localized bulging occurs. It is found from equation (\ref{eq2_12}) that $\lambda_b$ is a function of $\lambda_a$ and $\lambda_z$. According to the derivative of integral with variable bounds, we arrive at 
\begin{equation}
\left\{
\begin{aligned}
&\dfrac{\partial P^*}{\partial \lambda_a}=\int_{\lambda_b}^{\lambda_a}\dfrac{\partial P_1}{\partial \lambda_a}\textrm{d}\lambda+P_1(\lambda,\lambda_a,\lambda_z)\Big |_{\lambda=\lambda_a}-P_1(\lambda,\lambda_a,\lambda_z)\Big |_{\lambda=\lambda_b}\dfrac{\partial \lambda_b}{\partial \lambda_a}, \\&
\dfrac{\partial P^*}{\partial \lambda_z}=\int_{\lambda_b}^{\lambda_a}\dfrac{\partial P_1}{\partial \lambda_z}\textrm{d}\lambda-P_1(\lambda,\lambda_a,\lambda_z)\Big |_{\lambda=\lambda_b}\dfrac{\partial \lambda_b}{\partial \lambda_z},\\&
\dfrac{\partial N^*}{\partial \lambda_a}=\dfrac{\partial N_1}{\partial \lambda_a}\int_{\lambda_b}^{\lambda_a}N_2\textrm{d}\lambda+N_1\left(\int_{\lambda_b}^{\lambda_a}\dfrac{\partial N_2}{\partial \lambda_a}\textrm{d}\lambda+N_2(\lambda,\lambda_a,\lambda_z)\Big |_{\lambda=\lambda_a}-N_2(\lambda,\lambda_a,\lambda_z)\Big |_{\lambda=\lambda_b}\dfrac{\partial \lambda_b}{\partial \lambda_a}\right), \\&
\dfrac{\partial N^*}{\partial \lambda_z}=\dfrac{\partial N_1}{\partial \lambda_z}\int_{\lambda_b}^{\lambda_a}N_2\textrm{d}\lambda+N_1\left(\int_{\lambda_b}^{\lambda_a}\dfrac{\partial N_2}{\partial \lambda_z}\textrm{d}\lambda-N_2(\lambda,\lambda_a,\lambda_z)\Big |_{\lambda=\lambda_b}\dfrac{\partial \lambda_b}{\partial \lambda_z}\right).
\end{aligned}
\right.\label{eq3_4}
\end{equation}
Although all integrals in (\ref{eq3_4}) can not be solved analytically, we emphasize that it still formulates the theoretical foundation of our numerical scheme. If all geometric and material parameters are specified, we could numerically determine the implicit function of $\lambda_z(\lambda_a)$ according to $J(P^*, N^*)=0$. 

\begin{figure}[!htbp]
\centering
\subfigure[The resultant axial force is fixed.]{\includegraphics[scale=0.78]{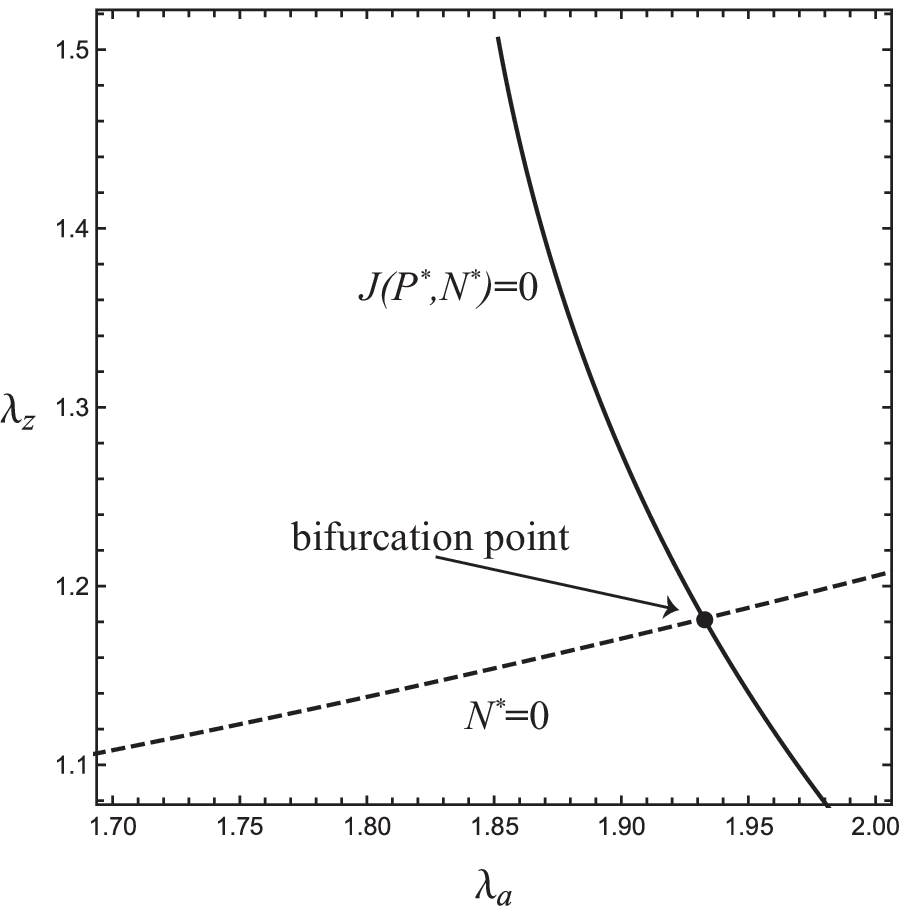}{\label{fig3a}}}\hspace{5mm}
\subfigure[The axial length is fixed.]{\includegraphics[scale=0.78]{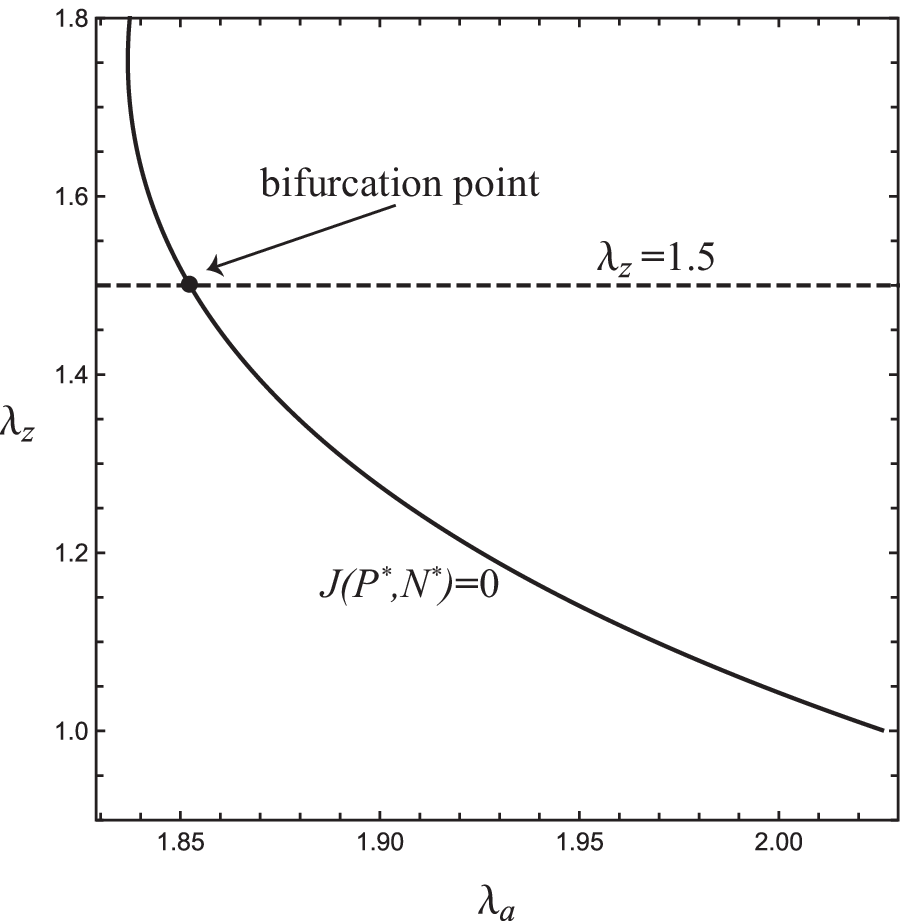}{\label{fig3b}}}
\caption{Contour plots of the bifurcation condition and the corresponding loading curves ($N^*=0$ or $\lambda_z=1.5$) when the exponential modulus gradient $(\ref{eq2_19})_2$ is used, and the parameters are given by $A^*=0.4$, $\beta_1=30$, $\zeta=30$ and $J_m=97.2$. The intersections indicate the first bifurcation point resulting in localized bulging.}\label{fig3}
\end{figure}

Next, we briefly summarize the solution procedure by specifying $A^*=0.4$, $\beta_1=30$, $\zeta=30$ and $J_m=97.2$. This parametric setting denotes a thick grade tube where the inner surface is stiffer. We first give an initial value $\lambda_z=1.001$ and seek a solution to $J(P^*, N^*)=0$ using Newton's iteration method. In doing so, a pair of solution $\{\lambda_z=1.001,\lambda_a=2.026\}$ can be obtained. We then adopt a loop algorithm starting from $\lambda_z=1.001$ with a small incremental step. It is well-known that the classical Newton's method for solving nonlinear algebraic equations is sensitive to the initial guess. Especially for complex problems, a proper initial guess can make the solution approach more efficient. Under the assumption that the curve of $J(P^*, N^*)=0$ is smooth, a nice initial guess can be found around the neighborhood of the solution of the previous step. By doing so, the relation between $\lambda_a$ and $\lambda_z$ can be found. We emphasize that this solution procedure can be applied to another nonlinear algebraic equation $N^*=0$, which offers the second implicit function of $\lambda_z$ and $\lambda_z$. For simplicity, extra computation details are omitted. In the case of fixed axial length, the value $\lambda_z=1.5$ supplies a horizontal line in the $(\lambda_a,\lambda_z)$-plane. We supply the graphical interpretation of the solution strategy for the bulge initiation. Currently, determination of the first bifurcation point is equivalent to identifying the intersection of $N^*=0$ and $J(P^*,N^*)=0$ or the counterpart of $\lambda_z=1.5$ and $J(P^*,N^*)=0$, as shown in Figure \ref{fig3}. For the prescribed parameters given earlier, we find that $\lambda_z^c=1.182$ and $\lambda_a^c=1.932$ if $N^*=0$ while $\lambda_a^c=1.852$ for $\lambda_z=1.5$.

So far, we have narrated the solution strategy for identifying the bulge initiation for graded soft tubes of arbitrary thickness. Especially, it is a numerical-scheme-based framework. Although we only employ the Gent model, this framework also supplies a calculation platform for a general material model and a generic modulus distribution. In the following analysis, we shall apply the solution procedure to the three different modulus functions in (\ref{eq2_19}) and aim to elucidate the influence of modulus gradient as well as the position where maximum modulus attains on the bulge initiation.
 
\subsection{LTI: Fixed axial force}
First of all, we consider the case of fixed axial force. The aim is to unravel the effect of different geometrical and material parameters and various modulus gradients on the bulge initiation. It is emphasized that there are at most four free parameters in the graded Gent model, videlicet, the dimensionless inner radius $A^*$, the ratio of shear modulus $\beta_1$ or $\beta_2$, the $J_m$ and the decay rate $\zeta$ in $(\ref{eq2_19})_2$. As introduced before, the last two parameters will be fixed by $J_m=97.2$ and $\zeta=30$ in all illustrative examples, and their influence is beyond the scope of this study. As a result, there are two independent parameters $A^*$ and $\beta_1$ (or $\beta_2$). The former measures the thickness of the graded tube while the latter displays the maximum modulus variation. In addition to these parameters, another important factor is the modulus distribution. Therefore, we shall assign a representative value of $\beta_1$ or $\beta_2$ and exhibit the results for different inner radii and repeat this procedure when $A^*$ turns into the fixed parameter.

\begin{figure}[!htbp]
\centering
\subfigure[$\beta_1=\beta_2=5$.]{\includegraphics[scale=0.78]{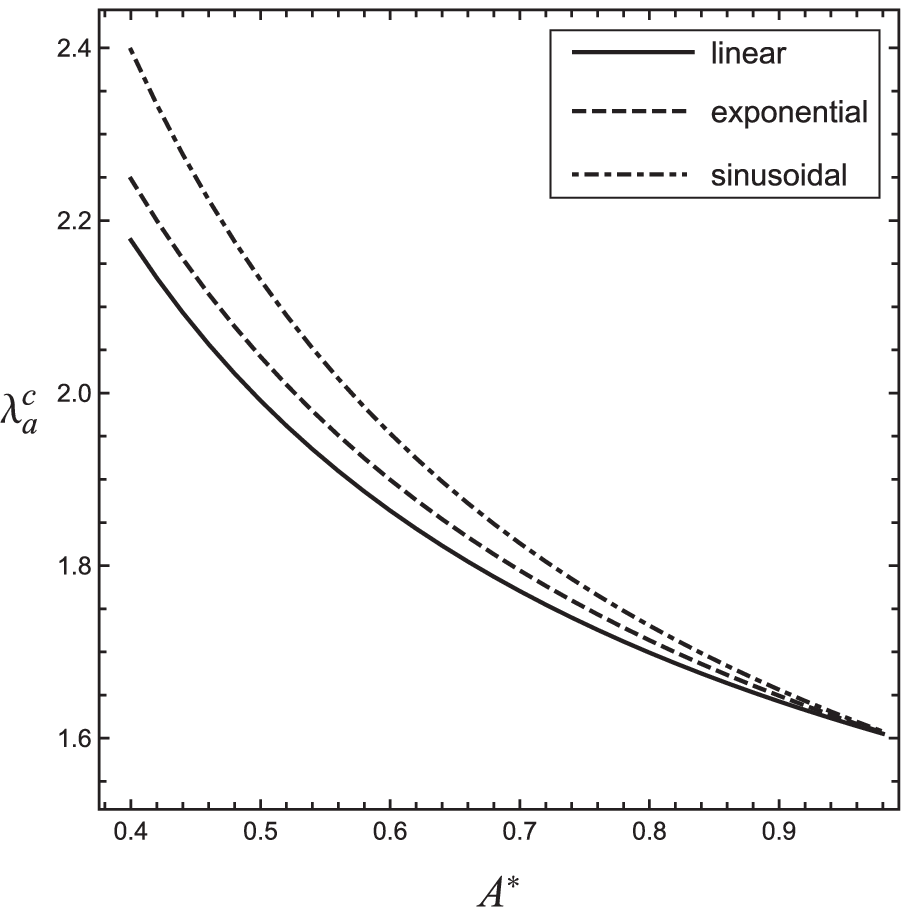}{\label{fig4a}}}\hspace{5mm}
\subfigure[$\beta_1=\beta_2=30$.]{\includegraphics[scale=0.78]{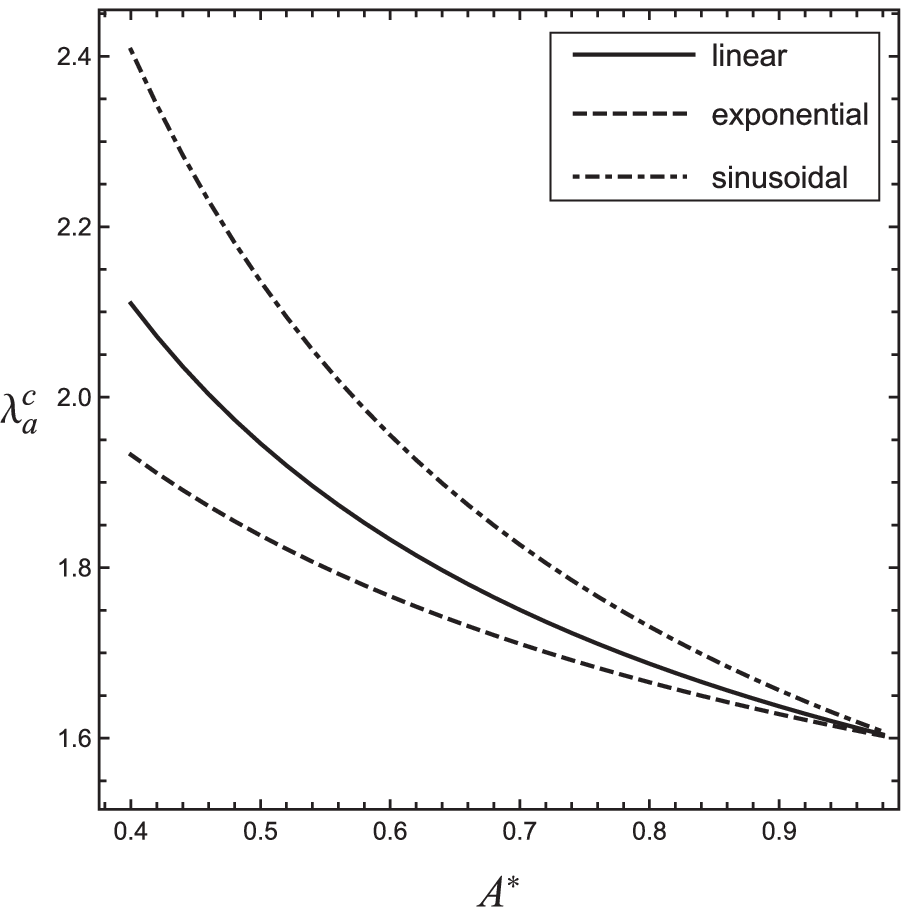}{\label{fig4b}}}
\caption{The critical stretch $\lambda_a^c$ versus the dimensionless inner radius $A^*$ when the three modulus functions in (\ref{eq2_19}) are adopted. The resultant axial force is given by $N^*=0$, and the other parameters are given by $\zeta=30$ and $J_m=97,2$.}\label{fig4}
\end{figure}

Applying the solution strategy outlined in the preceding subsection, the dependences of the critical stretch $\lambda_a^c$ on the dimensionless inner radius $A^*$ are shown in Figure \ref{fig4}. The modulus ratios are set to be $5$  in Figure \ref{fig4a} and $30$ in Figure \ref{fig4b}, respectively. It can be seen that the critical stretch $\lambda_a^c$ always decreases monotonically with increased $A^*$. Note that varying $A^*$ is equivalent to altering the thickness of the tube. This implies the fact that localized bulging is invariably easier to take place in a thinner tube, regardless of the modulus distribution. The curves for the sinusoidal function are always higher, leading to a greater critical stretch for a given $A^*$ and further a more stable structure compared to the other two modulus distributions. We observe that the curves for the sinusoidal function in both figures are almost identical. It is therefore concluded that the modulus ratio $\beta_2$ has a weak influence on the critical bifurcation stretch $\lambda_a^c$. Furthermore, the dashed line is higher than the solid one in \ref{fig4a} while becomes lower in \ref{fig4b}. This indicates that a graded tube with the shear modulus decaying exponentially from the inner surface is more stable than the counterpart where the shear modulus decays linearly when $\beta_1=5$. However, if $\beta_1=30$, a reverse inference can be obtained. Ultimately, as $A^*$ goes to unity, the deviations among all curves almost vanish, so modulus gradient can be ignored in extremely thin tubes.

\begin{figure}[!htbp]
\centering\includegraphics[scale=0.78]{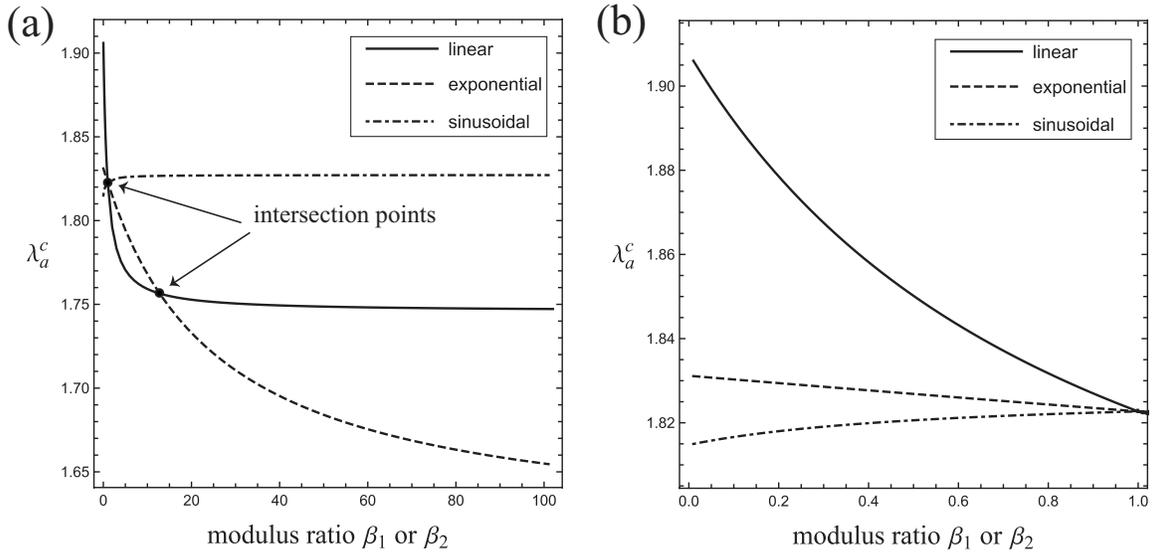}\caption{The critical stretch $\lambda_a^c$ versus the ratio of modulus when the three modulus functions in (\ref{eq2_19}) are adopted. (a) The modulus ratio ranges from 0.01 to 101 and (b) the modulus ratio ranges from 0.01 to 1. The right figure is a blowup of the right figure when the modulus ratio is small. The resultant axial force is given by $N^*=0$, and the other parameters are given by $A^*=0.7$, $\zeta=30$ and $J_m=97,2$.}\label{fig5}
\end{figure}

The above analysis primarily reveals the complexity of the influence of modulus gradient on the bifurcation threshold. In the following study, the modulus ratios $\beta_1$ and $\beta_2$ are taken into account to explore the impact of the modulus ratio as well as the modulus distribution. Figure \ref{fig5} displays the relations between $\lambda_a^c$ and the modulus ratio for three modulus gradients by specifying $A^*=0.7$. The right figure highlights the detail of the left one when the modulus ratio is around unity. Seen from Figure \ref{fig5}, we find that the critical stretch approaches a limit value for all modulus functions as $\beta_1$ or $\beta_2$ increases. If the shear modulus has a sinusoidal distribution, the bulge initiation is nearly independent of the modulus ratio $\beta_2$, especially when $\beta_2>1$. Yet the critical stretch is always a decreasing function for the other two modulus distributions. Specifically, if the shear modulus grows exponentially from the inner surface ($\beta_1<1$), the bifurcation threshold only varies in a very small range compared to the linearly increasing counterpart, and a linear distribution always delays the occurrence of localized bulging. Nevertheless, when $\beta_1>1$, the bifurcation threshold decreases fast and starts to be impervious to the variation of $\beta_1$ as $\beta_1$ approaches practically 30 when the linear function $(\ref{eq2_19})_1$ is selected.  On the one hand, it is observed from figure \ref{fig5} that these three curves connect at $\beta_1=\beta_2=1$ where a graded tube is reduced to a homogeneous one. On the other hand, the solid line and dashed line intersect at another point and we define the corresponding horizontal coordinate as $\beta_c$. It can be solved that $\beta_c=12.78$ for $A^*=0.7$. Accordingly, the exponential distribution retards the bifurcation for $1<\beta_1<\beta_c$ while diminishes the bulge initiation for $\beta>\beta_c$. In other words, compared to a linearly decayed shear modulus from the inner surface, an exponential counterpart gives rise to a more stable structure as $\beta$ is lower than $\beta_c$ while resulting in a less stable structure otherwise. 

\begin{figure}[!htbp]
\centering\includegraphics[scale=0.78]{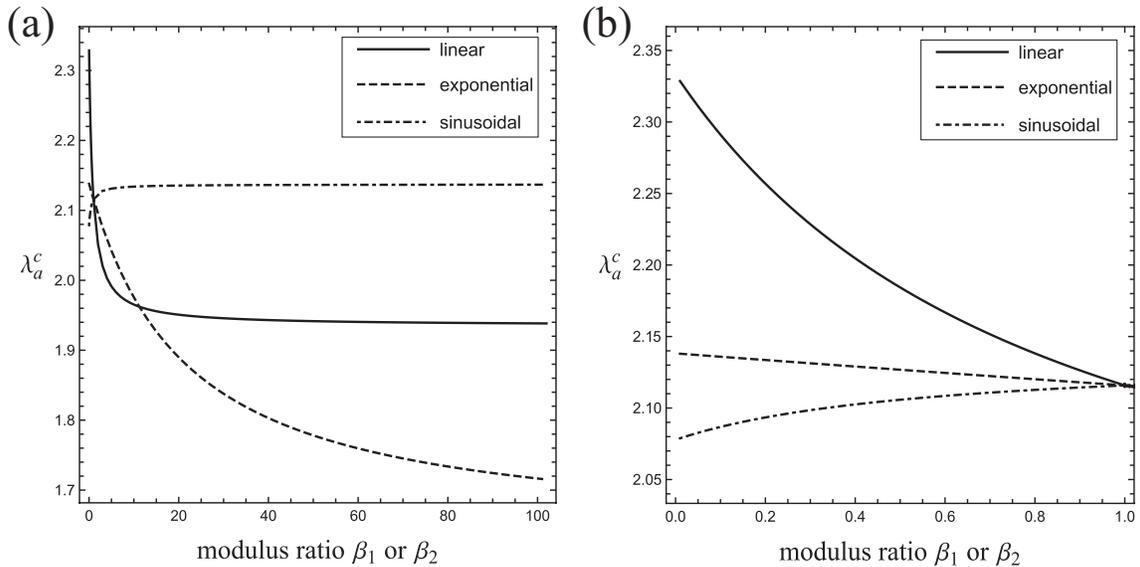}\caption{The critical stretch $\lambda_a^c$ versus the ratio of modulus when the three modulus functions in (\ref{eq2_19}) are adopted. (a) The modulus ratio ranges from 0.01 to 101 and (b) the modulus ratio ranges from 0.01 to 1. The right figure is a blowup of the right figure when the modulus ratio is small. The resultant axial force is given by $N^*=0$, and the other parameters are given by $A^*=0.5$, $\zeta=30$ and $J_m=97,2$.}\label{fig6}
\end{figure}

To verify that the main characteristic in Figure \ref{fig5} is irrelevant to the special choice $A^*=0.7$, we further present the dependence of $\lambda^c_a$ on the modulus ratio for three modulus functions in Figure \ref{fig6} by considering a thicker tube where $A^*=0.5$. Likewise, the right part exhibits a blowup when the modulus ratio is around unity. It can be seen that Figure \ref{fig6} differs from Figure \ref{fig5} in the range of the vertical axis. In addition to this small divergence, these two figures are qualitatively the same. So similar conclusion can be drawn and we identify $\beta_c=11.22$ for $A^*=0.5$. 

In summary, if other conditions remain identical and only the thickness can be varied, a higher thickness always delays bulge formation in a pressurized graded tube. However, the effect of the modulus gradient is quite distinctive. Although we only select three prototypical modulus functions as shown in (\ref{eq2_19}), the bifurcation threshold versus the modulus experiences various tendencies. When the shear modulus either grows monotonically or declines monotonically from the inner surface, the critical stretch $\lambda_a^c$ is constantly a decreasing function as $\beta_1$ increases and will attain a constant value for large enough $\beta_1$. This implies that the bifurcation threshold remains practically unaffected when the modulus mismatch between the inner and outer surfaces is fairly high. Furthermore, these two curves intersect at $\beta_1=1$ and $\beta_1=\beta_c>1$, leading to a transition zone $1<\beta_1<\beta_c$ where a graded tube with shear modulus distributing exponentially is more stable. Yet a non-monotonically sinusoidal distribution of the shear modulus leads to another situation. The curve of $\lambda_a^c$ versus $\beta_c$ is slowly increasing if $\beta_2<1$ and becomes approximately a horizontal line once $\beta_2$ passes the value of unity. Specifically, the critical stretch $\lambda_a^c$ is practically identical to that of the homogeneous counterpart. As a result, this special graded tube has similar deformation behavior to its homogeneous analogue. We mention that a graded tube with the shear modulus distributed sinusoidally can be viewed as an effectively continuous simulacrum of a sandwich tube. Thus, we note that a sandwich tube with a hard internal core can resist the internal pressure inducing aneurysm formation without influence on the critical stretch. From the viewpoint of a structure optimization, such a composition reduces material cost but increases the critical pressure giving rise to aneurysm formation. Bearing in mind that human arteries are composed of a sandwich structure where the core (intermediate layer) has the largest elastic modulus \citep{je2000,jctr2012}, our theoretical prediction agrees well with the natural evolution and optimization of human arteries.

\subsection{LTII: Fixed axial length}
Secondly, we address the other commonly adopted loading condition where the axial length is prescribed by stretching a rubber tube and then fixing two ends. A noteworthy feature in LTII is that the curve of pressure against stretch $\lambda_a$ or volume ratio $\lambda_a^2\lambda_z$ may lose $N$-shape and further may be a monotonically increasing function. Consequently, the limit-point theory fails to identify the bifurcation point, and we resort to the bifurcation condition derived by \cite{fu2016}. 

\begin{figure}[!htbp]
\centering
\subfigure[$\beta_1=\beta_2=5$.]{\includegraphics[scale=0.78]{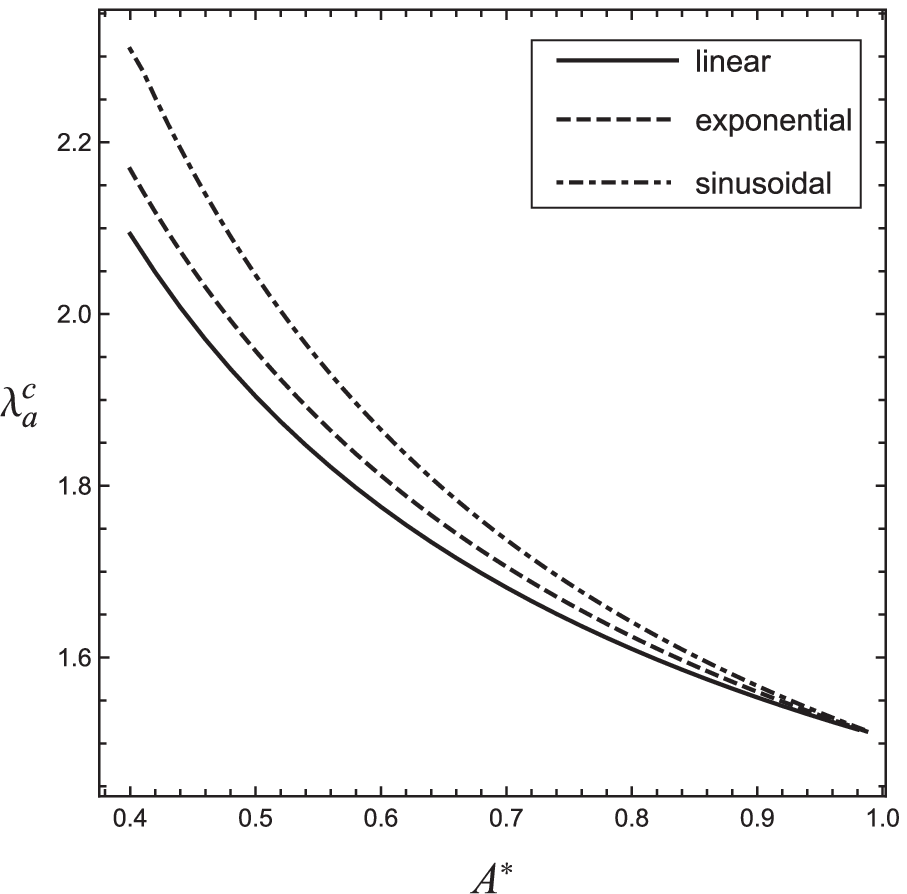}{\label{fig7a}}}\hspace{5mm}
\subfigure[$\beta_1=\beta_2=30$.]{\includegraphics[scale=0.78]{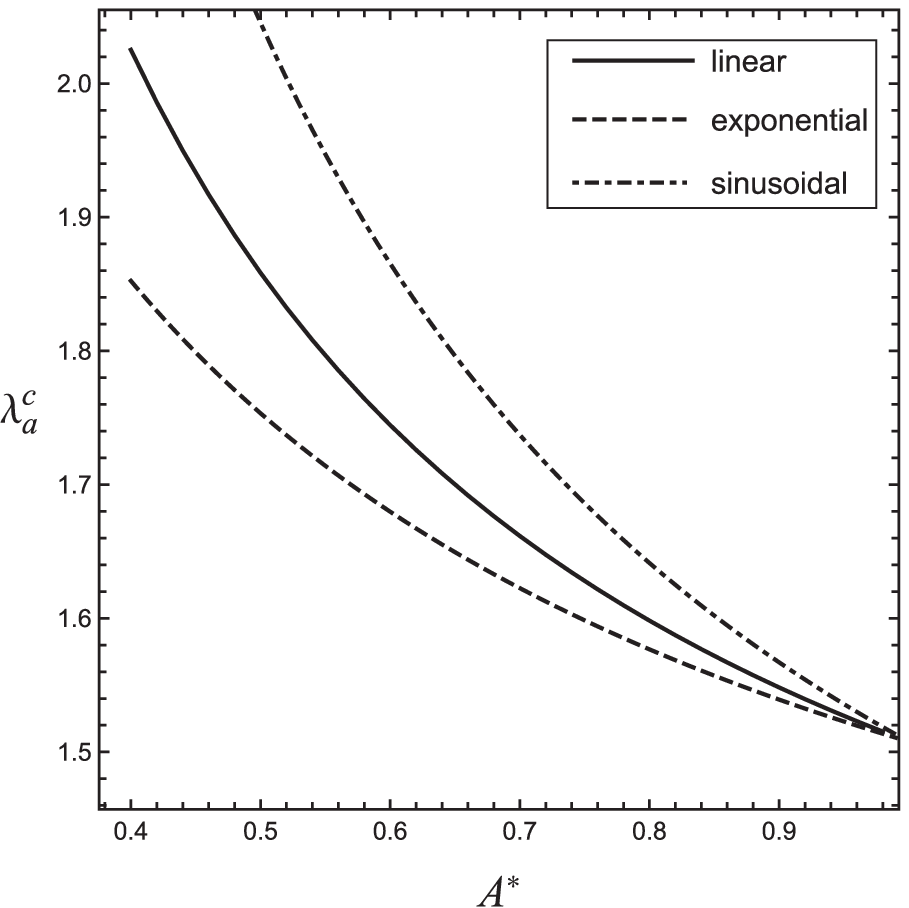}{\label{fig7b}}}
\caption{The critical stretch $\lambda_a^c$ versus the dimensionless inner radius $A^*$ when the three modulus functions in (\ref{eq2_19}) are adopted. The axial stretch is given by $\lambda_z=1.5$, and the other parameters are given by $\zeta=30$ and $J_m=97,2$.}\label{fig7}
\end{figure}

\begin{figure}[!htbp]
\centering\includegraphics[scale=0.78]{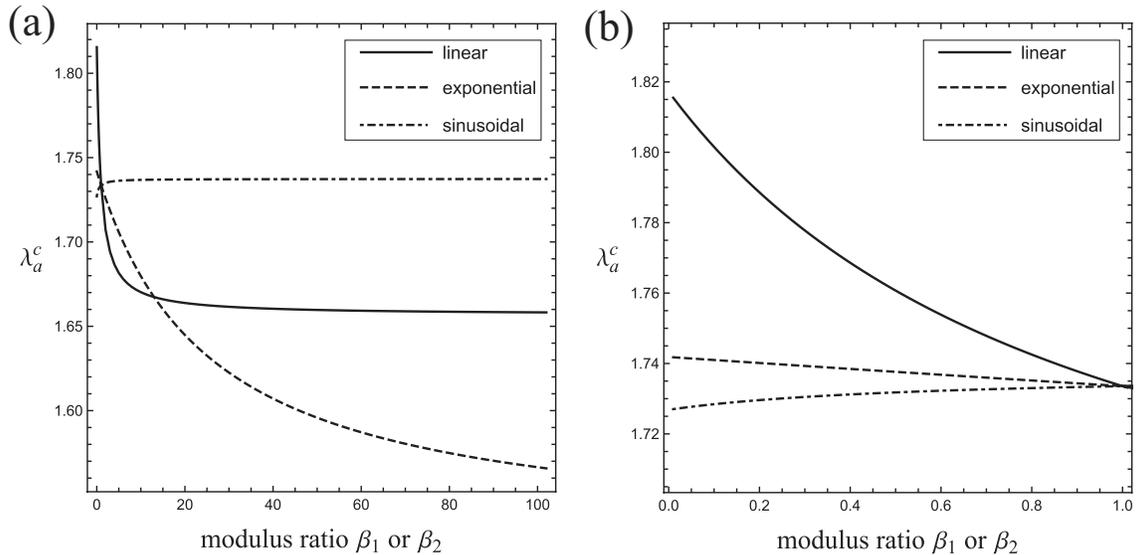}\caption{The critical stretch $\lambda_a^c$ versus the ratio of modulus when the three modulus functions in (\ref{eq2_19}) are adopted. (a) The modulus ratio ranges from 0.01 to 101 and (b) the modulus ratio ranges from 0.01 to 1. The right figure is a blowup of the right figure when the modulus ratio is small. The axial stretch is given by $\lambda_z=1.5$, and the other parameters are given by $A^*=0.7$, $\zeta=30$ and $J_m=97,2$.}\label{fig8}
\end{figure}

Previous studies of localized bulging in inflated bilayer tubes have indicated that the effect of material and geometrical parameters on the bifurcation threshold directing to localized bulging has no essential distinction \citep{liu2019,ye2019}. Therefore, we anticipate that similar bifurcation results as that in LTI will be observed here. To this end, we specify the modulus ratio and plot the dependence of $\lambda_a^c$ on the dimensionless inner radius $A^*$ in Figure \ref{fig7} while Figure \ref{fig8} displays $\lambda_a^c$ against the modulus ratio $\beta_1$ or $\beta_2$. As explained earlier, the axial stretch is given by $\lambda_z=1.5$ in all illustrative examples. For comparison, Figure \ref{fig7a} shows the outcomes for three modulus functions when $\beta_1=\beta_2=5$ while Figure \ref{fig7b} exhibits the counterparts when $\beta_1=\beta_2=30$. As expected, the curve for the sinusoidal function is the highest. In addition, the curve for a linear function is lower than the one for exponential function as $\beta_1=5$ but becomes higher when $\beta_1$ is transferred to 30. These features are consistent with those in Figure \ref{fig4}, so we omit more detailed descriptions.

Then we explore the influence of modulus ratio depicted in Figure \ref{fig8}. It can be seen that the tendencies and shapes of all curves are consistent with those in Figures \ref{fig5} and \ref{fig6}. Similarly, we determine the second intersect of the solid line and the dashed line as $\beta_c=13.01$. 

At present, the influence of the tube thickness, the shear modulus ratio, and the modulus distribution on bulge initiation is revealed by the use of the bifurcation condition $J(P^*, N^*)=0$ and based on the solution procedure established at the beginning of this section. We emphasize that in practical applications not only the bulge formation but also the bulge evolution is of great interest because material rapture often occurs in the growth or propagation stage. However, theoretically characterizing the solution path for bulge growth is extremely difficult. Only weakly nonlinear analysis can be carried out for thick tubes in the framework of finite elasticity \citep{ye2020}. Unfortunately, the information of weakly nonlinear behavior has limited guidance in the bulge growth stage since the bifurcation nature of localized bulging is subcritical, and experimentally observed amplitude is always finite. To acquire more knowledge of bulge growth in inflated graded tubes, we shall establish a finite element model in commercial software Abaqus in the next section.

\section{Finite element model}
In this section, we pursue constructing a finite element (FE) model in Abaqus to conduct a post-buckling analysis of localized bulging in a pressurized graded tube. Although the incompressible Gent model has not been built in the software, one could write UHYPER subroutine codes following the user guideline \citep{abaqus}. In our previous studies \citep{liu2019,ye2019}, homogeneous Gent model was supplemented into Abaqus employing UHYPER subroutine coding. Another important issue is to model modulus gradient. In practice, we can discretize the tube into many sub-tubes in the thickness direction and each sub-tube is composed of a homogeneous Gent material where the shear modulus can be determined according to the given modulus function. This is equivalent to approximating a continuous modulus function by a staircase one. As the mesh size tends to zero, the deformation behavior of a layered tube would perfectly coincide with that of the original graded one. Based on this idea, pattern transition induced by volumetric growth in graded structures was investigated using FE analysis based on this idea \cite{liu2020}. Moreover, another more convenient way is to make use of the linearly temperature-dependent elastic modulus option in Abaqus \citep{prsa2014,chen2018}. In this alternative approach, the temperature distribution will exactly conform to the modulus distribution and the thermal expansion coefficient must be set to be zero. 

In principle, both approaches can be applied to simulate the deformation and instability of a graded tube under the combined action of internal pressure and axial extension in Abaqus. Notwithstanding, it should be pointed out that the former method may need more grids to guarantee the quality of meshes and may further require additional computation costs. So we employ the latter method and write the UHYPER subroutine codes in Fortran for the graded Gent model by creating a connection between the shear modulus and the temperature field. Ultimately, a graded Gent model with arbitrary modulus function is well established in Abaqus. In doing so, we can carry out FE simulations for the localized bulging.

Before proceeding further, we shall check the validity of the FE model. For that purpose, we calculate the bifurcation threshold using FE analysis and compare the results with the corresponding theoretical predictions. To eliminate the end effects of a grade tube with finite length, we construct all FE models by specifying the length to overall diameter by 30, or equivalently $L/B=60$. Furthermore, the built-in command ``Buckle" can not be applied to solve the eigenvalue problem when the critical mode is zero, and we utilize the Module ``Static, Riks" to perform a fully nonlinear analysis. 

To trigger localized bulging, certain structures or physical imperfections need to be involved in the FE model. In the case of fixed axial force, we refer to the practical end conditions in an inflation experiment where both the two ends are fastened and the radial movement is confined \citep{wang2019}. So in the FE model, the radial displacements on the two ends are restricted and this can be regarded as a geometrical imperfection. For fixed axial length, the shear modulus at the center $Z=0$ is assigned to $\mu(R)-0.001$ if $\beta_1$ (or $\beta_2$) is less than 30 or otherwise a greater magnitude $\mu(R)-0.01$ shall be used. In addition, A  quadratic 3D hybrid element with reduced integration (C3D20RH) is employed in all subsequent FE simulations.

\begin{figure}[!htbp]
\centering
\subfigure[The modulus ratio is given by $\beta_1=30$.]{\includegraphics[scale=0.78]{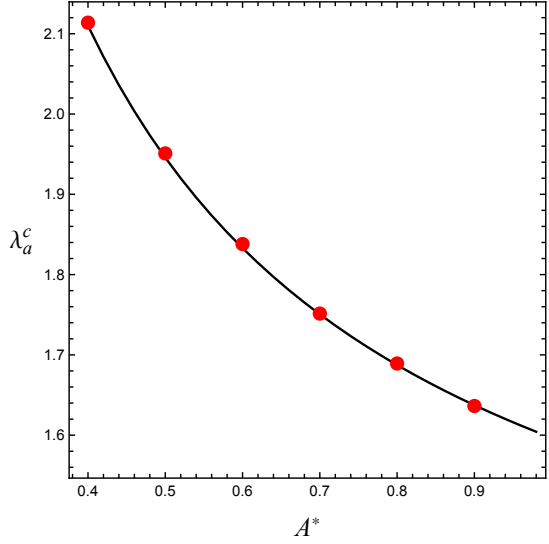}{\label{fig9a}}}\hspace{5mm}
\subfigure[The inner radius is given by $A^*=0.7$.]{\includegraphics[scale=0.78]{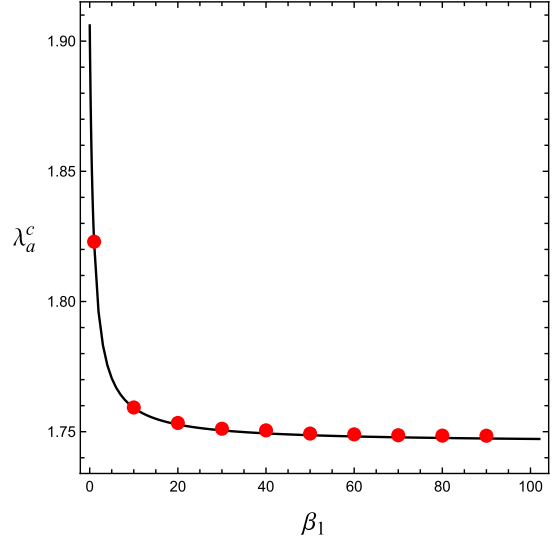}{\label{fig9b}}}
\subfigure[The modulus ratio is given by $\beta_1=30$.]{\includegraphics[scale=0.78]{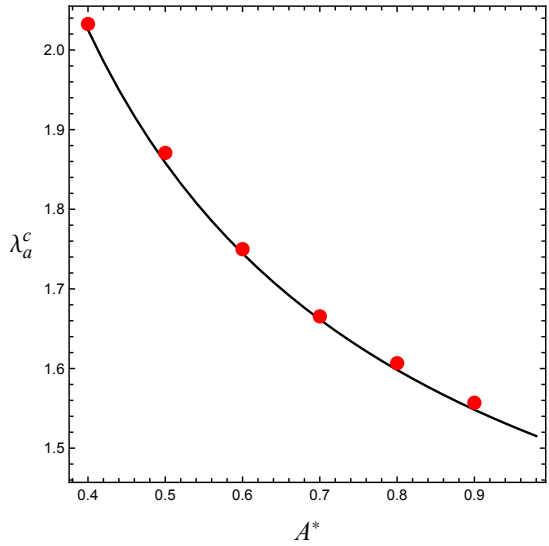}{\label{fig9c}}}\hspace{5mm}
\subfigure[The inner radius is given by $A^*=0.7$.]{\includegraphics[scale=0.78]{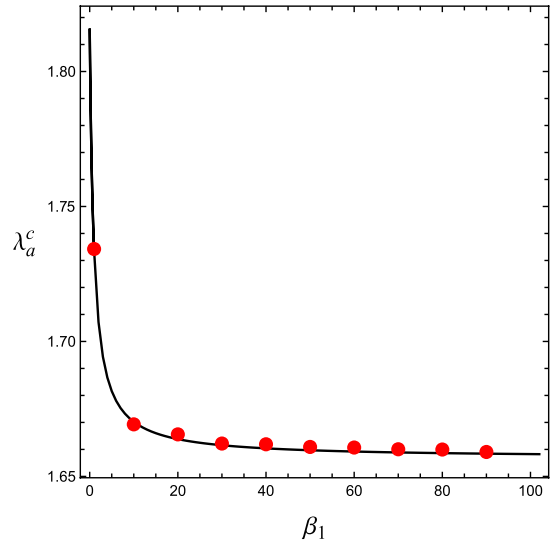}{\label{fig9d}}}
\caption{(Color online) Comparisons of the critical stretch $\lambda_a^c$ between the theoretical predictions and the FE results when the shear modulus varies linearly. The resultant axial force is fixed by $N^*=0$ in top subfigures while the axial stretch is given by $\lambda_z=1.5$ in the bottom subfigures. The red dots denote the FE solutions and the solid lines correspond to the theoretical ones.}\label{fig9}
\end{figure}

For convenience, the FE model is validated by taking the linear modulus function $(\ref{eq2_19})_1$ as an example, and all comparisons are summarized in Figure \ref{fig9}. The theoretical solutions are denoted by solid lines while the FE results are represented by red dots. The resultant axial force is specified by $N^*=0$ in Figures \ref{fig9a} and \ref{fig9b} and the axial stretch is fixed by $\lambda_z=1.5$ in Figures \ref{fig9c} and \ref{fig9d}, respectively. It is found that the critical stretch based on FEA agrees well with that from the theoretical model. This implies the credibility and robustness of the established FE model and the UHYPER subroutine codes for the graded Gent model.

Now, a robust FE model for simulating localized bulging in pressurized graded tubes has been formulated. In particular, the modulus gradient in the FE model can be arbitrary. So the FE model paves a convenient way to implement a fully nonlinear analysis of bulge evolution. In the next section, both bulge growth and bulge propagation will be investigated for a graded tube.

\section{Bulge growth and propagation}
It is known that a complete process of localized bulging in pressurized tubes contains three typical stages, consisting of bulge initiation, bulge growth, and bulge propagation. When the internal pressure passes a critical value, a bulge profile appears. With increased pressure, the diameter of the bulge grows until it reaches a maximum. Then we define a new parameter $\lambda_a^m$ that depicts the largest hoop stretch at the inner surface. As a result, the $\lambda_a^mA^*$ indicates the maximum inner radius for the bulge. Afterwards, the growing bulge ceases to expand in the radial direction and only dilates in the axial direction at constant pressure, and the instability enters into the propagation stage. We emphasize that an ideal propagation stage only exists in an infinitely long tube, so the maximum radius cannot be exactly reached in a tube of finite length.

In many actual applications, such as soft robots and flexible actuation, bulge profile is harmful to the normal function of these devices, so the parametric analysis for the onset of localized bulging can supply useful insight into bulge suppression. If localized bulging has emerged, material rupture may appear either in the growth stage or during the period of bulge propagation. In practice, the rupture risk is possibly relevant to the size of a bulge or the magnitude of circumferential stretch. For clinicians, the diameter of an aneurysm can be used to evaluate the danger of aneurysm fracture \citep{jb2012}. It is therefore of great significance to estimate how large a bulge can finally attain. As noted earlier, human arteries are composed of three layers where the intermediate layer has the highest elastic modulus, and we can approximately take a non-monotonic modulus distribution to qualitatively elucidate the mechanism behind artery evolution. Hence, the section aims to reveal the effect of modulus ratio and material gradient on the maximum magnitude of the bulge and to provide further insight into how a normal artery is constructed to resist aneurysm formation while keeping proper elasticity.

\begin{figure}[!htbp]
\centering\includegraphics[scale=1]{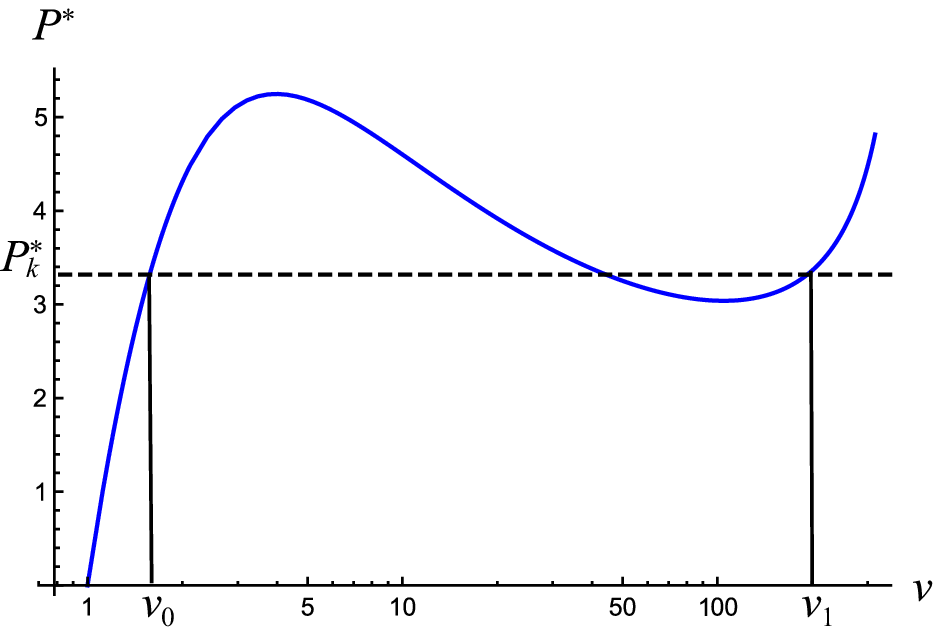}\caption{(Color online) The dependence of pressure $P^*$ on volume ratio $v$ for the primary deformation subjected to inflation of a graded tube where the sinusoidal distribution $(\ref{eq2_19})_3$ is applied. The resultant axial force is fixed by $N^*=0$ and the parameters are give by $A^*=0.7$ and $\beta_2=30$. }\label{fig10}
\end{figure}

It is pointed out that the propagation stage is a two-phase deformation and is composed of two uniform states joined by a transition zone. When the resultant axial force is fixed, the curve of pressure versus $v$ or $\lambda_a$ presents an $N$-shape where the first stationary point corresponds to the onset of localized bulging. The $v=\lambda_a^2\lambda_z$ denotes the ratio of the volume enclosed in the undeformed tube to that in the inflated state. In this scenario, the pressure under which a bulge starts to propagate can be determined according to Maxwell's equal-area rule \citep{jam1984}. To clearly illustrate this methodology, we consider a graded tube with sinusoidally varying shear modulus subjected to inflation, and the resultant axial force is fixed by $N^*=0$. Figure \ref{fig10} sketches the relation between the dimensionless pressure $P^*$ and the volume ratio $v$ when the dimensionless inner radius $A^*$ is set to be 0.7. The marked pressure $P_k^*$ depicts the origination of bulge propagation. Maxwell's equal-area rule yields
\begin{align}
P^*_k(v_1-v_0)=\int_{v_0}^{v_1}P^*(v)\textrm{d}v, \label{eq5_1}
\end{align}
where $v_0$ corresponds to a state before localized bulging and $v_1$ the counterpart after bulge has propagated two both ends. Referring to the solution strategy outlined in Section 3.1, we can also program in $Mathematica$ to identify the exact values of $P_k^*$, $v_0$ and $v_1$, and the technicality is neglected for brevity. For instance, for the parameters used in plotting Figure \ref{fig10}, we acquire that $P_k^*=3.318$, $v_0=1.575$ and $v_1=192.87$. However, when the axial length is specified, the pressure versus $v$ or $\lambda_a$ may become monotonic, so Maxwell's equal-area rule is no longer valid. Notwithstanding, it is naturally expected that the principal conclusion for fixed axial force can offer qualitatively guidance for other loading methods as usual.

\begin{figure}[!htbp]
\centering\includegraphics[scale=1]{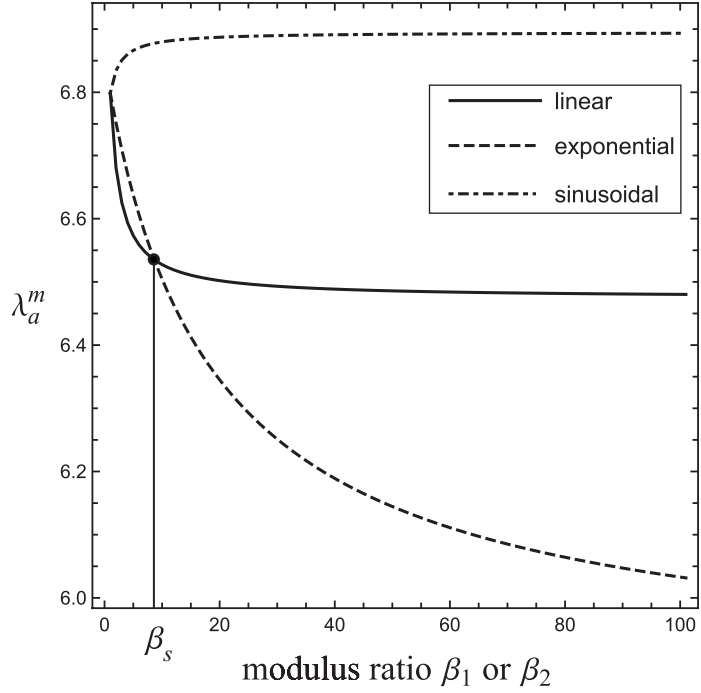}\caption{Illustration of dependences of $\lambda_a^m$ on $\beta_1$ when the linear and exponential functions are employed or $\beta_2$ when the sinusoidal function is applied. The resultant axial force is given by $N^*=0$ and the dimensionless inner radius $A^*$ is prescribed by $0.7$.}\label{fig11}
\end{figure}
 
In the subsequent analysis, we shall illuminate the influence of the modulus ratio as well as the material gradient on the maximum circumferential stretch $\lambda_a^m$ based on equation (\ref{eq5_1}). Similar to the parametric analysis for the onset of localized bulging, we also plot in Figure \ref{fig11} the curves for three modulus functions in (\ref{eq2_19}) when the modulus ratio $\beta_1$ or $\beta_2$ ranges from 1 to 101 and the dimensionless inner radius is given by $A^*=0.7$. This implies that the shear modulus of the inner (or middle) surface is equal to or larger than that of the outer surface. Moreover, it is a wonder that the trend of each curve is identical to that in Figures \ref{fig5} and \ref{fig6}. Similarly, the sinusoidal distribution is highest among these curves, indicating that a localized bugle can attain a greater radius in this case compared to a linearly decayed or an exponentially decayed modulus distribution if the peak value of shear modulus and the thickness remain the same. Meanwhile, the curve of sinusoidal distribution increases slightly from $\beta_2=1$ and soon becomes virtually flat as $\beta_2$ is around 10. Consequently, such a special non-monotonic modulus gradient brings a negligible change in the final deformed state. Note that these curves intersect at $\beta_1=\beta_2=1$ where a homogeneous counterpart is obtained. Furthermore, we only focus on the linear and exponential distributions since the unique difference is the modulus gradient. It can be seen that both curves are monotonically decreasing as $\beta_1>1$. Especially, within the domain $1<\beta_1<\beta_s$ where $\beta_s=8.67$, a pressurized graded tube with the shear modulus distributed exponentially can finally undergo a larger bulge compared to the linear counterpart if the axial force is specified. On the other hand, with increased $\beta_1$, the dependence of the final magnitude of a bulge on the modulus ratio becomes weaker and weaker until the modulus ratio has no influence.

\begin{figure}[!htbp]
\centering\includegraphics[scale=1]{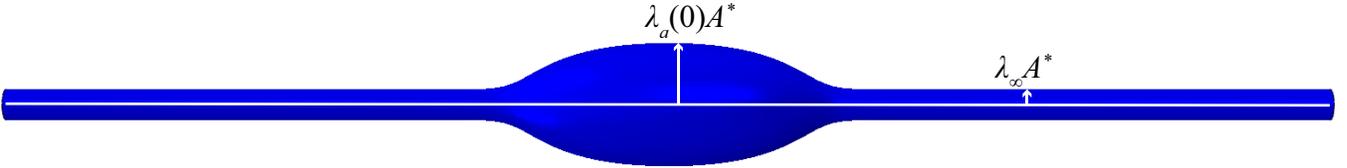}\caption{(Color online) Graphical interpretation of the scaled amplitude $\lambda_a(0)-\lambda_\infty$. The white horizontal line marks the axis of the tube.}\label{fig12}
\end{figure}

As of now, we have performed a theoretical analysis on the limit magnitude of the bulge in a graded tube subject to combined fixed axial force and internal pressure in the light of Maxwell's equal area. An interesting discovery is that a sinusoidal modulus function does practically not alter the final size of a bulge compared to its homogeneous counterpart. Bearing in mind that a similar situation takes place for the bulge initiation, we are left to speculate what will happen in the bulge growth stage for a homogeneous tube and a graded tube where the shear modulus distributes sinusoidally. To this end, we apply the established FE model in the previous section and carry out a fully nonlinear analysis. At first, we define the amplitude of localized bulging. As illustrated in Figure \ref{fig12}, the hoop stretch is dependent not only on the radial coordinate $R$ but also on the axial coordinate $Z$ after bulge initiation. In a perfect system, namely, the tube is infinitely long, bulge solution holds translation invariance. So we can assume that localized bulging, associated with the greatest diameter of the bulge, always appears at the center of the tube $Z=0$. Then the dimensionless radius at the inner surface and $Z=0$ writes $\lambda_a(0)A^*$. Furthermore, we denote the corresponding radius for the uniform inflation region by $\lambda_\infty A^*$, where $\lambda_\infty$ stands for the hoop stretch at the infinity. In a uniform inflation process, the $\lambda_a$ is independent of $Z$ hence we get $\lambda_a=\lambda_\infty$. In an illustrative FE example, we identify $\lambda_\infty$ in the non-bulging region, as shown in Figure \ref{fig12}. The scaled amplitude of a bulge is given by $\lambda_a(0)-\lambda_\infty$. 

\begin{figure}[!htbp]
\centering
\subfigure[The resultant axial force is fixed by $N^*=0$.]{\includegraphics[scale=0.9]{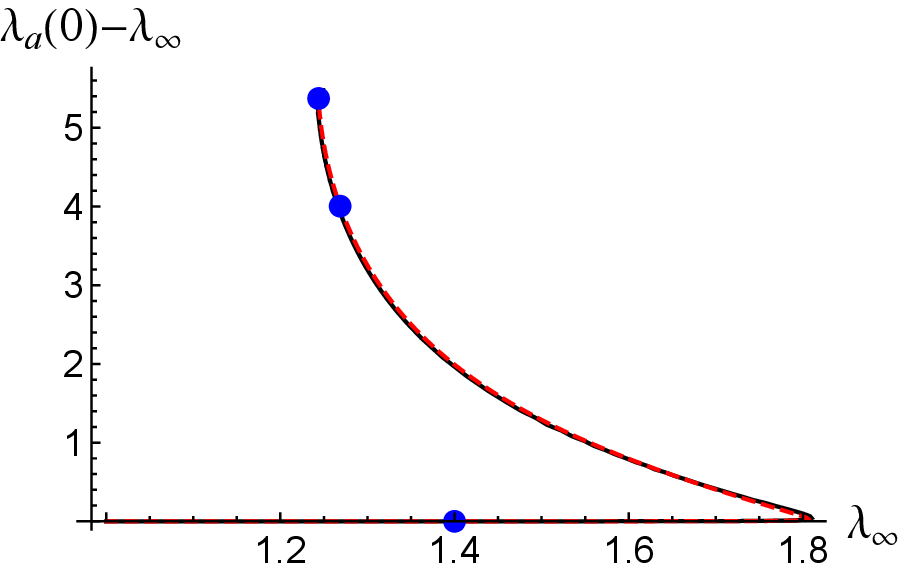}{\label{fig13a}}}\hspace{5mm}
\subfigure[The axial length is fixed by $\lambda_z=1.5$.]{\includegraphics[scale=0.9]{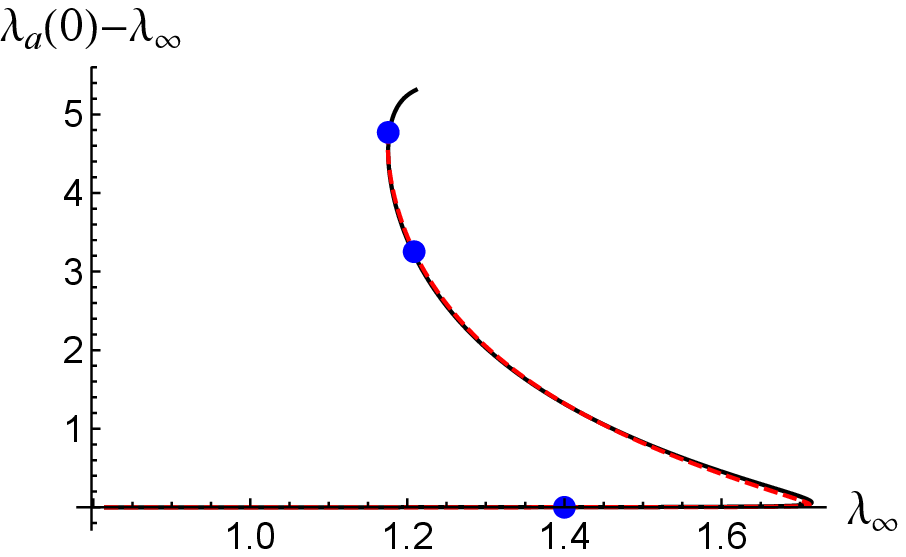}{\label{fig13b}}}
\caption{(Color online) Comparisons of the bifurcation diagrams between a graded tube where the shear modulus function is given by $(\ref{eq2_19})_3$ and a homogeneous counterpart. The parameters are given by $A^*=0.7$ and $\beta_2=30$. The black solid lines represent the results for the homogeneous tube while the red dashed lines the counterparts for the graded one. The profiles of these blue dots highlighted in the figure are displayed in Figures \ref{fig14} and \ref{fig15}, respectively.}\label{fig13}
\end{figure}

On specifying $A^*=0.7$, we show the bifurcation diagrams for a homogeneous tube and a graded one in Figure \ref{fig13} using FE analysis. The modulus ratio for the graded tube is given by $\beta_2=30$, and the $\beta_2$ is reduced to unity in a homogeneous tube. In particular, Figure \ref{fig13a} displays the results when the resultant axial force is prescribed while Figure \ref{fig13b} plots the counterparts when the axial length is fixed. From the bifurcation diagram, it is observed that the amplitude of bulge is trivial in the uniform inflation where $\lambda_\infty$ is reduced to the circumferential stretch in the primary deformation $\lambda_a$. As the inflation continues, especially when $\lambda_\infty$ passes the critical value $\lambda_a^c$, a non-trivial solution resulting in localized bulging takes place and the scaled amplitude $\lambda_a(0)-\lambda_\infty$ gets a positive value. Afterwards, the amplitude increases with reduced $\lambda_\infty$. This implies the fact that the existing bulge dilates in both the radial and the axial directions while the non-bulging part shrinks. Remarkably, we again find an extremely good coincidence for the deformation processes between the homogeneous tube and the graded one no matter it is the resultant axial force $N^*$ or the axial stretch $\lambda_z$ that is fixed. Combining the analysis for the bulge initiation and the maximum size of a bulge, it is therefore summarized that a graded tube with sinusoidally distributed shear modulus can undergo a higher internal pressure and remain the deformation feature at the same time. For instance, the deformation paths in Figure \ref{fig13a} are identical while the associated critical pressures for a graded tube and a homogeneous one are $P_{cr}^*=5.248$ and $P_{cr}^*=0.271$, respectively. 

\begin{figure}[!htbp]
\centering\includegraphics[scale=0.8]{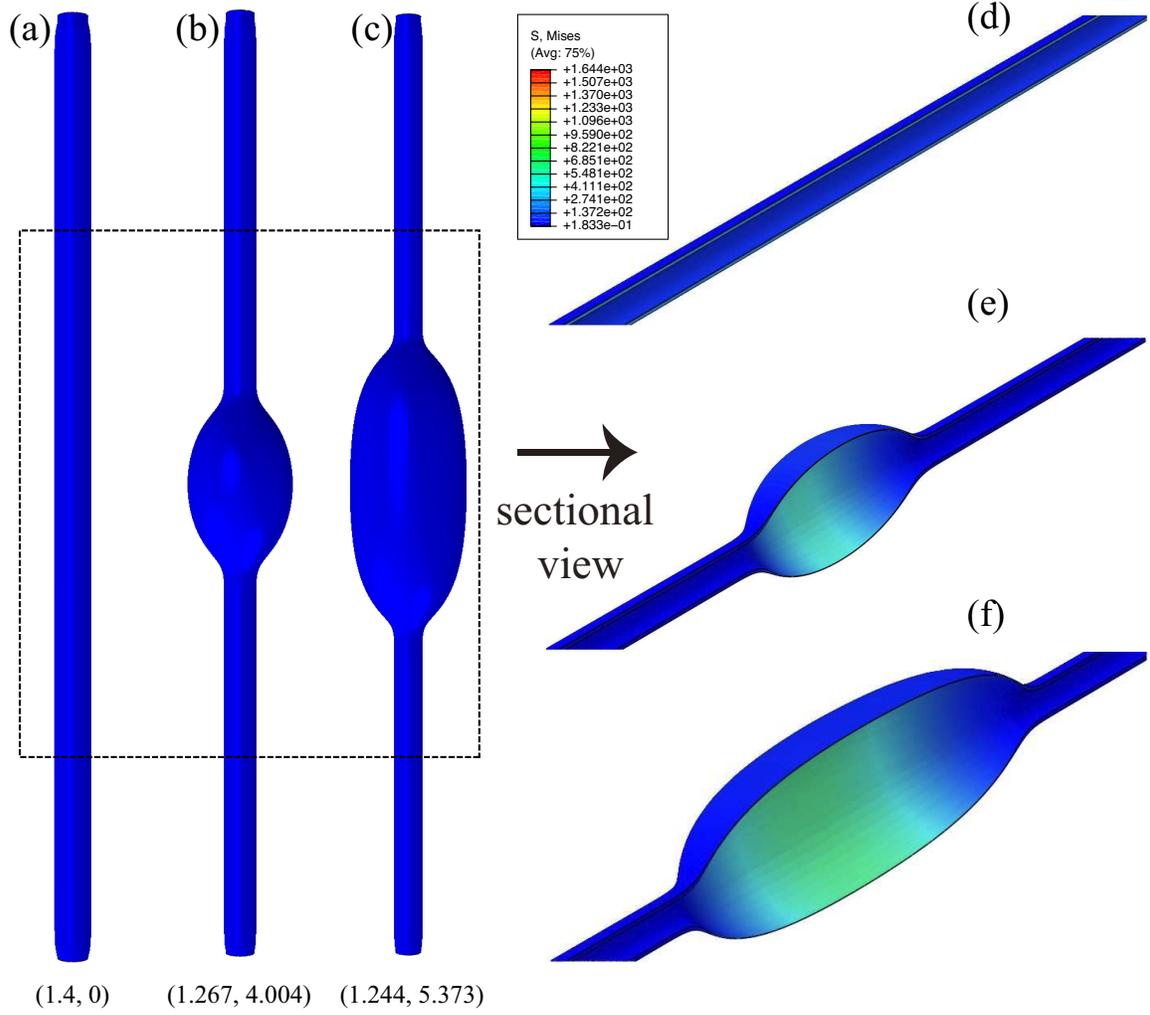}\caption{(Color online) Deformed profiles of the three marked points in Figure \ref{fig13a} for the graded tube when the axial force is fixed by $N^*=0$. (a) to (c) show the entire deformations at different periods while (d) to (e) illustrate the sectional views. The values of $\lambda_\infty$ and $\lambda_a(0)-\lambda_\infty$ are listed below the corresponding subfigures. The parameters are given by $A^*=0.7$ and $\beta_2=30$.}\label{fig14}
\end{figure}

We also highlight three pairs of $\{\lambda_\infty,\lambda_a(0)-\lambda_\infty\}$ by blue dots for the graded tube in Figures \ref{fig13a} and \ref{fig13b}, respectively. In particular, we select a state before instability and two states after bulge initiation. The corresponding coordinates $\{\lambda_\infty,\lambda_a(0)-\lambda_\infty\}$ read $\{1.4,0\}$, $\{1.267, 4.004\}$ and $\{1.244, 5.373\}$ from the bottom to the top in Figure \ref{fig13a} while in Figure \ref{fig13b} the counterparts are given by $\{1.4,0\}$, $\{1.209, 3.253\}$ and $\{1.176, 4.771\}$. The deformed profiles for the three highlighted points are exhibited in Figure \ref{fig13a}. In the right part, we further plot the cross-sections enclosed in the dashed rectangular to offer a clear view of the stress distribution on the inner surface. The aspect ratio of length to diameter is 30. As introduced in Section 4, the radial displacement of two ends are restricted and this setup serves as a geometric imperfection that can be used to trigger aneurysm formation at a critical pressure. It can be seen that the end effect decays very fast and has no influence on the deformation around the center where localized bulging is expected to appear. As the stretch passes the critical stretch $\lambda_a^c$, a bulge occurs at the center of the tube and starts to expand in the radial and axial orientations. At the same time, the inner or outer diameter at the non-bulging area (outside the dashed rectangular in Figure \ref{fig14}) suffers a small but continuous drop, which explains why the curve in the amplitude diagram turns back when $\lambda_\infty$ exceeds the critical stretch $\lambda_a^c$. From the sectional view, it is also found that the stress and strain are both concentrated on the center of the bulge, which may cause potential rapture. 

\begin{figure}[!htbp]
\centering\includegraphics[scale=0.8]{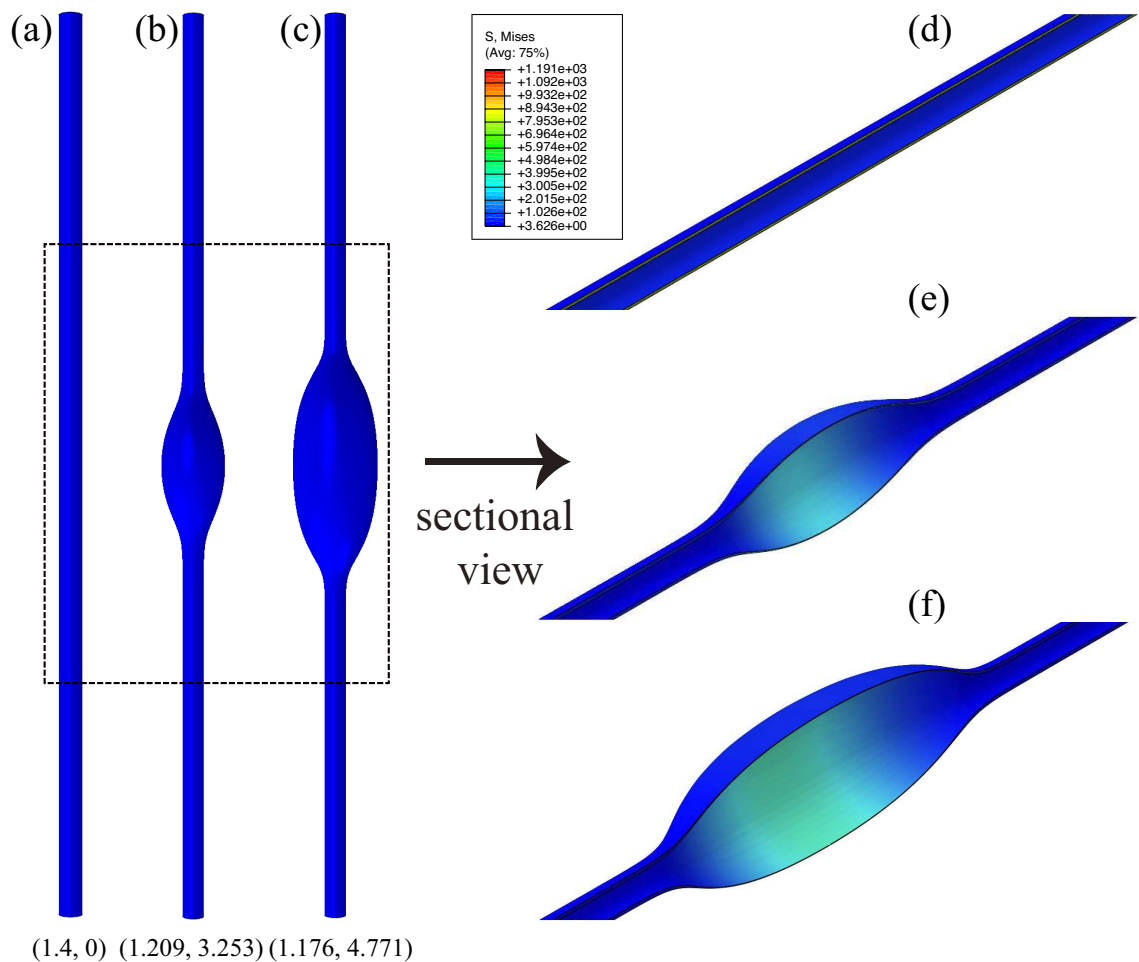}\caption{(Color online) Deformed profiles of the three marked points in Figure \ref{fig13b} for the graded tube when the axial stretch is fixed by $\lambda_z=1.5$. (a) to (c) show the entire deformations at different periods while (d) to (e) illustrate the sectional views. The values of $\lambda_\infty$ and $\lambda_a(0)-\lambda_\infty$ are listed below the corresponding subfigures. The parameters are given by $A^*=0.7$ and $\beta_2=30$.}\label{fig15}
\end{figure}

Finally, we plot in Figure \ref{fig15} the deformed configurations of all blue points in Figure \ref{fig13b}. The layout is similar to that in Figure \ref{fig14}. We emphasize that an alternative physical defect that the shear modulus at the middle cross-section $Z=0$ is reduced to $\mu(R)-0.01$ is employed in the case of fixed axial length. In doing so, the center of the tube is slightly softer than other positions. Then localized bulging occurs at the center if the internal pressure attains a critical value. Additionally, the cross-sections in the dashed rectangular are shown to depict evidently the stress distribution on the inner surface.

\section{Concluding remarks}
Localized bulging or aneurysm formation in inflated graded cylindrical tubes of arbitrary thickness was investigated in detail within the framework of finite elasticity. The primary deformation where a tube dilates uniformly was analytically determined to formulate the expressions of the internal pressure and the resultant axial force. Meanwhile, a concise bifurcation condition in terms of the internal pressure and the resultant axial force has been applied and a semi-analytical solution procedure was proposed to determine the onset of localized bulging. Two typical loading conditions, defined by fixed axial force and fixed axial length, were adopted throughout the paper. In particular, the case of fixed axial force has potential applications in soft pneumatic actuators \citep{he2020,li2022} while the other end condition corresponds to the \textit{in vivo} status of human arteries \citep{bmm2014}. For all illustrative examples, the incompressible Gent material was engaged and three archetypal modulus gradients were used, containing a linear, an exponential, and a sinusoidal function. Then a thorough theoretical analysis was conducted for the critical stretch $\lambda_a^c$ at the inner surface. Note that the structure is more stable with a higher $\lambda_a^c$ in a volume control problem. In this sense, a thicker tube is always less susceptible to suffering localized bulging regardless of a concrete form of the material gradient. Remarkably, it turns out that the critical stretch is sensitive to the specific modulus distribution applied. In particular, a sinusoidal function almost does not influence $\lambda_a^c$, no matter what values of the ratio of the shear modulus on the middle position to that on the outer surface (denoted by $\beta_2$). Furthermore, a linear or an exponential one will reduce $\lambda_a^c$ as the ratio of the shear modulus on the inner surface to that on the outer surface (denoted by $\beta_1$) increases. However, a larger enough $\beta_1$ ceases to affect the critical stretch. It is emphasized that a qualitatively similar conclusion can be obtained for both end conditions. The current analysis on bulge initiation implies that not only the modulus mismatch but also the exact position where maximum modulus attains is vital to the initiation of localized bulging. Bearing in mind that an accurate measurement of elastic modulus of human arteries is extremely difficult \citep{prsa2019} and we only have a general understanding that the media (intermediate layer) of an artery occupies a larger elastic modulus \cite{je2000,jctr2012}, the current analysis concerning sinusoidally distributed shear modulus implies that the deformation is almost insensitive to the practical value of the shear modulus of the middle surface. 

To track the evolution of a bulge, a finite element model incorporating material inhomogeneity is established using Abaqus UHYPER subroutine coding. This model can be used to perform nonlinear analysis for localized bulging in inflated graded tubes for an arbitrary modulus gradient. The robustness of the established FE model was validated by comparing the bulge initiations using finite element analysis to the theoretical predictions when a linear modulus gradient is applied. Ultimately, bulge propagation for the case of fixed axial force was analytically studied according to Maxwell's equal-area rule, and the influence of material gradient, as well as the modulus ratio on the maximum magnitude of a bulge, were clarified. Combing the analysis for the onset of localized bulging, it is found that either the material gradient or the modulus ratio has the same effect on the deformed profiles at the critical state and at the propagation stage. Furthermore, we compared the bifurcation diagrams between a graded tube where the shear modulus decays sinusoidally from the middle surface to both lateral boundaries and its homogeneous counterpart based on finite element simulations. Interestingly, these two bifurcation diagrams are nearly identical for both loading types. Therefore, this work provides an answer for one fundamental problem arising from structure optimization, i.e. stiffening the intermediate surface enhances the ability of a structure to resist internal pressure, and neither the critical stretch nor the deformation path varies compared with the homogeneous counterpart. This conclusion is perfectly consistent with the actual situation of human arteries. It is expected that the current analysis would supply useful insight into localization instabilities in graded structure as well as into aneurysm prevention in functional soft devices.

\section*{Acknowledgments}
The work was supported by the National Natural Science Foundation of China (Grant Nos 12072227, 12072225, and 12121002). The Abaqus simulations were carried out on TianHe-1 (A) at the National Supercomputer Center in Tianjin, China. We thank Prof. Yibin Fu at Keele University for valuable advice and discussion.

\end{document}